\documentclass[reprint,superscriptaddress,nofootinbib,eqsecnum,amsmath,amssymb,aps,pra]{revtex4-2}

\usepackage{amsmath}
\usepackage{graphicx}
\usepackage{color}
\usepackage{hyperref}
\usepackage{ulem}

\begin{document}

\title{Spin vs. position conjugation in quantum simulations with atoms:\\ application to quantum chemistry}%

\author{N.A. Moroz}
\affiliation{Quantum Technology Centre, M.V.~Lomonosov Moscow State University\\ Leninskiye Gory 1-35, 119991, Moscow, Russia}
\author{K.S. Tikhonov}%
\affiliation{St. Petersburg State University, 199034, St. Petersburg, Russia}%
\affiliation{Russian Quantum Center, Skolkovo, Moscow 143025, Russia}
\author{L.V. Gerasimov}
\affiliation{Quantum Technology Centre, M.V.~Lomonosov Moscow State University\\ Leninskiye Gory 1-35, 119991, Moscow, Russia}
\affiliation{Centre for Interdisciplinary Basic Research, HSE University, St. Petersburg 190008, Russia}
\author{A.D. Manukhova}
\affiliation{Department of Optics, Palack\'{y} University, 17 Listopadu 12, 771 46 Olomouc, Czech Republic}
\author{\\I.B. Bobrov}
\affiliation{Quantum Technology Centre, M.V.~Lomonosov Moscow State University\\ Leninskiye Gory 1-35, 119991, Moscow, Russia}
\affiliation{Russian Quantum Center, Skolkovo, Moscow 143025, Russia}
\author{S.S. Straupe}
\affiliation{Quantum Technology Centre, M.V.~Lomonosov Moscow State University\\ Leninskiye Gory 1-35, 119991, Moscow, Russia}%
\affiliation{Russian Quantum Center, Skolkovo, Moscow 143025, Russia}
\author{D.V. Kupriyanov}\email{kupriyanov@quantum.msu.ru}
\affiliation{Quantum Technology Centre, M.V.~Lomonosov Moscow State University\\ Leninskiye Gory 1-35, 119991, Moscow, Russia}
\affiliation{Centre for Interdisciplinary Basic Research, HSE University, St. Petersburg 190008, Russia}
\affiliation{Department of Physics, Old Dominion University\\ 4600 Elkhorn Avenue, Norfolk, Virginia 23529, USA}

\date{\today}%

\begin{abstract}
\noindent The permutation symmetry is a fundamental attribute of the collective wavefunction of indistinguishable particles. It makes a difference for the behavior of collective systems having different quantum statistics but existing in the same environment. Here we show that for some specific quantum conjugation between the spin and spatial degrees of freedom the indistinguishable particles can behave similarly for either quantum statistics. In particular, a mesoscopically scaled collection of atomic qubits, mediated by optical tweezers, can model the behavior of a valent electronic shell compounded with nuclear centers in molecules. This makes possible quantum simulations of mono and divalent bonds in quantum chemistry by manipulation of up to four bosonic atoms confined with optical microtraps. 
\end{abstract}

\maketitle


\section{Introduction}
\noindent Recent progress in the preparation of atomic lattices has shown them as a promising platform for programmable quantum simulations of discretized systems in statistical physics with a large number of atomic qubits involved \cite{Lukin2017,Lukin2021}. That concerns general issues of the phase transition theory and, in particular, has allowed observation of such subtle effects as non-adiabatic dynamics in topological phase conversion \cite{Lukin2019} and spatial transfer of the entanglement of atomic qubits \cite{Lukin2022}. A fault-tolerant multi-qubit quantum computer on neutral atoms can be attainable in the future \cite{Radnaev2024,Reichardt2024}, but its successful development could be promoted by important physical implications. 

The problem of describing the behaviour of electron subsystems in complex organic molecules might seem as such a possible implication if we could map their equilibrium states or dynamical evolution onto a multi-qubit quantum register. Indeed, although many approximation methods have been introduced in quantum chemistry, the complexity of quantum mechanics remains hard to simulate by purely classical algorithms. That is why quantum chemistry is under close attention for quantum computational protocols based on second quantisation and algorithm of the variational quantum eigensolver \cite{Cao2019}. Nevertheless an ideal quantum mapping of a system state and precise protocol execution would be difficult to do with currently existing experimental capabilities and still needs support from classical supercomputers \cite{Hempel2018,Motta2023,Robledo-Moreno2024,Greenediniz2024}. 

The concept of universal and highly scalable quantum computations with atomic spin qubits is not so easy to realize, and in the case of operational logic gates, commonly based on the Rydberg blockade protocol, there are various non-negligible and non-excluded physical disturbances preventing ideal quantum data processing \cite{Gerasimov2022,Vybornyi2024}. Some state-of-the-art optimization protocols indicate a potentially attainable fidelity for entangling operations up to $F\sim 99\%$ \cite{Saffman2023,Radnaev2024} and additional non-trivial ways of fidelity improvement have been proposed in \cite{Saffman2024a,Saffman2024b}. 

At present the realization of a scalable set of pre-tolerant quantum logic operations seems quite a challenging task, and the problem can be reevaluated in a more practical way towards so-called "noisy intermediate-scale quantum" (NISQ) devices, the perspectives of which are nicely reviewed in \cite{Zoller2022}. An example of noise tolerant analogue quantum simulation of the Born-Oppenheimer Hamiltonian of an electronic subsystem by emulation with an array of fermionic atoms has been proposed in \cite{Cirac2019,Arguello_Luengo2022}. The NISQ concept seems closely connected with proposals on observations and practical implementation of matter wave Hong-Ou-Mandel interference of ultracold atoms  \cite{Kaufman2015,Brandt2018,Regal2018,Aspect2020}, and phase transitions and self-organization in degenerate quantum systems \cite{Ritsch2017,Ritsch2019}. All these open optimistic perspectives towards developing various physical schemes of quantum simulations with low sensitivity to environmental disturbances.

In this paper, we show that the existing and already well-established experimental technique of spatial control of ultracold atoms, confined with optical microtraps, can be suggested for relevant analogue quantum simulations of covalent bonding in chemistry without extra technical amplifications. In our proposal we focus on matter wave interference in reproducing the electronic charge distribution in a molecule as a main effect. More concretely, the atoms, each loaded in its individual trap, play the role of electrons, and the optical trap potentials emulate the electron interaction with the nuclei. 

We show that the bosonic nature of atoms does not prevent a realistic visualization of a fermionic system consisting of up to four interacting electrons. Our proposal is crucially based on a unique property of the Young diagrams describing the collective states with minimal total spin angular momentum: in the case of three and four identical particles the symmetric and antisymmetric quantum states, associated with and constructed by these diagrams, have equivalent regular and transposed representations for both quantum statistics. In conventional chemistry, supporting the concept of covalent bonding, just such spin states, having closed antiferromagnetic electronic configurations, play a dominant role in stabilization of the compound system built up by positive and negative charges.\\%

The paper is organized as follows: In Section \ref{Section_II} we review the Hong-Ou-Mandel effect for bosonic and fermionic quantum statistics, highlighting the principle similarities in its manifestation. Then in Section \ref{Section_III} we give an explanation of this effect for atomic bosons having one-half pseudospin and confined with dipole microtraps. Section \ref{Section_IV} discusses the spin and position conjugation in systems of three and four indistinguishable particles. The respective analysis of the symmetric groups is given in Appendix \ref{Appendix_A} and the orbital representation of the wavefunction is given in Appendix \ref{Appendix_B}. Finally we present the results of our numerical simulations of the joint probability distribution, bunching and anti-bunching effects for the atomic matter waves and their dependence on the mutual configuration of the microtrap sites, which realises 
the similar behavior to electronic charge distributions in molecular divalent bonds. As an important consequence, which we discuss, the considered systems show double degeneracy of their ground states i.e. a matter wave constructed qubit structure. 

\section{Bunching and anti-bunching of two indistinguishable particles}\label{Section_II}

\noindent In this section, we overview the well-known Hong-Ou-Mandel interference effect, which we link with (i) the second quantized formalism and (ii) the concept of quantum scattering theory. That lets us clarify both difference and similarity in quantum conjugation between spin and position variables for two different quantum statistics. 

In accordance with the general principles of scattering theory, the free dynamics of the incoming wave packet, interacting with a scattering sample, is violated and the scattering $S$-matrix transforms the state of the system from an infinite past $|\psi\rangle_{\mathrm{in}}$ to an infinite future $|\psi\rangle_{\mathrm{out}}$ as a result of the interaction process \cite{GoldWat1964}. In the interaction representation, the corresponding asymptotic transformation is given by
\begin{equation}
|\psi\rangle_{\mathrm{out}}=\mathrm{e}^{\frac{i}{2\hbar}\hat{H}_0\tau}\mathrm{e}^{-\frac{i}{\hbar}\hat{H}\,\tau}\mathrm{e}^{\frac{i}{2\hbar}\hat{H}_0\tau}|\psi\rangle_{\mathrm{in}}\equiv%
\hat{S}|\psi\rangle_{\mathrm{in}},
\label{2.1}
\end{equation}
where $\tau\to+\infty$, $\hat{H}$ is the system Hamiltonian and $\hat{H}_0$ is its non-interacting part. The operator $\hat{S}$ can be represented as a matrix in a decoupled basis of two interacting subsystems, which we specify as $|\phi_i\rangle$ for the initial and $|\phi_{f}\rangle$ for the final states of the system.

Here we are interested in the case when the initial state consists of two separated particles which can interact only via the externally controllable macroscopic object, such as beamsplitter, waveguide, trap potential, etc. 
Let us assume that originally the particles income in two channels, specified as $1$ and $2$, 
and in two either polarization states, specified as $H$ (horizontal) and $V$ (vertical), 
or spin states, specified as $H$ (directed along a horizontal quantization axis) and $\bar{H}$ (oppositely directed). 
Applying the second quantized description we get
\begin{equation}
|\psi\rangle_{\mathrm{out}}=\hat{S}a^{\dagger}_{1X}a^{\dagger}_{2Y}|0\rangle,
\label{2.2}
\end{equation}
where $a^{\dagger}_{1X}$ and $a^{\dagger}_{2Y}$ are the creation operators, acting on the vacuum state $|0\rangle$ with $X$ and $Y$ each running either $H,V$ or $H,\bar{H}$. This transformation can be equivalently written as
\begin{equation}
|\psi\rangle_{\mathrm{out}}=\hat{S}a^{\dagger}_{1X}\hat{S}^{-1}\hat{S}a^{\dagger}_{2Y}\hat{S}^{-1}\hat{S}|0\rangle,
\label{2.3}
\end{equation}
and $\hat{S}$ is a unitary operator obeying $\hat{S}^{-1}=\hat{S}^{\dagger}$.

We assume the scattering object to be a linear device providing only coherent mixing of the original channels such that
\begin{eqnarray}
\hat{S}|0\rangle&=&\mathrm{e}^{i\alpha_{\Sigma}}|0\rangle%
\nonumber\\%
\nonumber\\%
\hat{S}a^{\dagger}_{1X}\hat{S}^{-1}&=&\makebox{$\frac{1}{\sqrt{2}}$}\,b^{\dagger}_{1X}+\makebox{$\frac{1}{\sqrt{2}}$}\,b^{\dagger}_{2X}%
\nonumber\\%
\hat{S}a^{\dagger}_{2Y}\hat{S}^{-1}&=&\makebox{$\frac{1}{\sqrt{2}}$}\,b^{\dagger}_{1Y}-\makebox{$\frac{1}{\sqrt{2}}$}\,b^{\dagger}_{2Y}%
\label{2.4}%
\end{eqnarray}
which implies that vacuum is unchanged and can only accumulate a certain phase shift $\alpha_{\Sigma}$ in all the modes associated with the outgoing scattering channels.\footnote{This extra phase is usually omitted, being included in general phase uncertainty of a vacuum state.} Two other equations express the intrinsically classical legacy in transformation of the field operators. The transformed basic field operators of the input channels (creators for $|\phi_i\rangle$-state with $i=1X,2Y$) are expressed by a linear combination of similar operators in the output channel (creators for $|\phi_f\rangle$-states with $f=1X,2Y$). In optics, the latter property is fundamentally based on the Fresnel reflection principles applied to an ideal transparent glass plate adjusted only for the coherent surface scattering.

We have used here the real + symmetric + Hermitian = unitary $2\times 2$ transformation matrix in (\ref{2.4}). 
Such a choice seems quite unique, but it is convenient, often applied, and, in fact, has no lack of generality. Let us follow how this transformation works for different quantum statistics.

\subsubsection{Bose-Einstein statistics}

\noindent 
The creation and annihilation operators, belonging to different modes, commute before and after the scattering event. If two quasiparticles (photons) arrive in equal polarizations we obtain
\begin{eqnarray}
|\psi\rangle_{\mathrm{out}}&=&\hat{S}a^{\dagger}_{1X}a^{\dagger}_{2X}|0\rangle
=\frac{1}{2}\left(b^{\dagger}_{1X}+b^{\dagger}_{2X}\right)\left(b^{\dagger}_{1X}-b^{\dagger}_{2X}\right)|0\rangle%
\nonumber\\%
&=&\frac{1}{2}\left[b^{\dagger}_{1X}b^{\dagger}_{1X}-b^{\dagger}_{2X}b^{\dagger}_{2X}\right]|0\rangle%
=\frac{1}{\sqrt{2}}\left[|2_{1X}^{\phantom{\dagger}}\rangle-|2_{2X}\rangle\right]%
\nonumber\\%
\label{2.5}
\end{eqnarray}
or explicitly
\begin{eqnarray}
|1_{1H},1_{2H}\rangle&\to&\frac{1}{\sqrt{2}}\left[|2_{1H},0_{2}^{\phantom{\dagger}}\rangle-|0_{1},2_{2H}\rangle\right]%
\nonumber\\%
|1_{1V},1_{2V}\rangle&\to&\frac{1}{\sqrt{2}}\left[|2_{1V},0_{2}^{\phantom{\dagger}}\rangle-|0_{1},2_{2V}\rangle\right]%
\label{2.6}%
\end{eqnarray}
which reveal an example of so called $NOON$-states i.e. bunching of the photons superposed between the spatially separated channels.

If two photons arrive in different polarizations symmetrically shared between the channels
\begin{equation}
|\psi\rangle_{\mathrm{in}}=\frac{1}{\sqrt{2}}\left[|1_{1H},1_{2V}^{\phantom{\dagger}}\rangle+|1_{1V},1_{2H}\rangle\right]%
\label{2.7}
\end{equation}
we obtain
\begin{eqnarray}
|\psi\rangle_{\mathrm{out}}&=&\hat{S}\frac{1}{\sqrt{2}}\left[a^{\dagger}_{1H}a^{\dagger}_{2V}+a^{\dagger}_{1V}a^{\dagger}_{2H}\right]|0\rangle%
\nonumber\\%
&=&\frac{1}{2\sqrt{2}}\left[\left(b^{\dagger}_{1H}+b^{\dagger}_{2H}\right)\left(b^{\dagger}_{1V}-b^{\dagger}_{2V}\right)\right.%
\nonumber\\%
&&\left.+\left(b^{\dagger}_{1V}+b^{\dagger}_{2V}\right)\left(b^{\dagger}_{1H}-b^{\dagger}_{2H}\right)\right]|0\rangle%
\nonumber\\%
&=&\frac{1}{\sqrt{2}}\left[b^{\dagger}_{1H}b^{\dagger}_{1V}-b^{\dagger}_{2H}b^{\dagger}_{2V}\right]|0\rangle%
\nonumber\\%
&=&\frac{1}{\sqrt{2}}\left[|1_{1H},1_{1V}^{\phantom{\dagger}}\rangle-|1_{2H},1_{2V}\rangle\right]%
\label{2.8}
\end{eqnarray}
which again reveal a bunched outcome
\begin{eqnarray}
\lefteqn{\frac{1}{\sqrt{2}}\left[|1_{1H},1_{2V}^{\phantom{\dagger}}\rangle+|1_{1V},1_{2H}\rangle\right]}%
\nonumber\\%
&\to&\frac{1}{\sqrt{2}}\left[|1_{1H},1_{1V}^{\phantom{\dagger}}\rangle-|1_{2H},1_{2V}\rangle\right]%
\label{2.9}%
\end{eqnarray}
The main difference with (\ref{2.6}) is that the paired photons appear in both the output channels in different polarizations.

It makes difference if the photons arrive in different polarizations arranged as anti-symmetric state
\begin{equation}
|\psi\rangle_{\mathrm{in}}=\frac{1}{\sqrt{2}}\left[|1_{1H},1_{2V}^{\phantom{\dagger}}\rangle-|1_{1V},1_{2H}\rangle\right]%
\label{2.10}
\end{equation}
Then we obtain
\begin{eqnarray}
|\psi\rangle_{\mathrm{out}}&=&\hat{S}\frac{1}{\sqrt{2}}\left[a^{\dagger}_{1H}a^{\dagger}_{2V}-a^{\dagger}_{1V}a^{\dagger}_{2H}\right]|0\rangle%
\nonumber\\%
&=&\frac{1}{2\sqrt{2}}\left[\left(b^{\dagger}_{1H}+b^{\dagger}_{2H}\right)\left(b^{\dagger}_{1V}-b^{\dagger}_{2V}\right)\right.%
\nonumber\\%
&&\left.-\left(b^{\dagger}_{1V}+b^{\dagger}_{2V}\right)\left(b^{\dagger}_{1H}-b^{\dagger}_{2H}\right)\right]|0\rangle%
\nonumber\\%
&=&\frac{1}{\sqrt{2}}\left[b^{\dagger}_{2H}b^{\dagger}_{1V}-b^{\dagger}_{2V}b^{\dagger}_{1H}\right]|0\rangle%
\nonumber\\%
&=&-\frac{1}{\sqrt{2}}\left[|1_{1H},1_{2V}^{\phantom{\dagger}}\rangle-|1_{1V},1_{2H}\rangle\right]%
\label{2.11}
\end{eqnarray}
Thus $|\psi\rangle_{\mathrm{out}}=-|\psi\rangle_{\mathrm{in}}$ i.e. such state is an eigenstate of the beamsplitter.

\subsubsection{Fermi-Dirac statistics}

\noindent The specific of the second quantized description for fermions dictates that any collective quantum state would be defined as occupation of an ordered orbital set. 
Let us explain this using an example of an arbitrary basis state
\begin{equation}
|\psi\rangle=|n_0,n_1,\ldots n_i,\ldots n_k,\ldots\rangle%
\label{2.12}
\end{equation}
where the individual orbital sites are enumerated by integers $0,1\ldots i, \ldots k, \ldots$. The state is actually represented by the Slater determinant constructed by the occupied sites, and it depends on the order in sequence of the site indices. 
Normally, the natural order $0<1<\ldots <i<\ldots <k$ is assumed.

Then the basic annihilation and creation second quantized operators are defined as
\begin{eqnarray}
a_i|n_0,n_1\ldots 1_i\ldots n_k\ldots\rangle&=&(-)^{\Sigma(0,i)}|n_0,n_1\ldots 0_i,\ldots n_k,\ldots\rangle%
\nonumber\\%
a_i|n_0,n_1\ldots 0_i\ldots n_k\ldots\rangle&=&0\cdot |n_0,n_1\ldots 0_i,\ldots n_k,\ldots\rangle%
\nonumber\\%
\nonumber\\%
a_i^{\dagger}|n_0,n_1\ldots 0_i\ldots n_k\ldots\rangle&=&(-)^{\Sigma(0,i)}|n_0,n_1\ldots 1_i,\ldots n_k,\ldots\rangle%
\nonumber\\%
a_i^{\dagger}|n_0,n_1\ldots 1_i\ldots n_k\ldots\rangle&=&0\cdot |n_0,n_1\ldots 1_i,\ldots n_k,\ldots\rangle%
\nonumber\\%
\label{2.13}
\end{eqnarray}
where $\Sigma(0,i)=\sum_{j=0}^{i-1}n_j$ is the sum of the occupation numbers for $j<k$.

As a particular example, if $i<k$ we obtain
\begin{eqnarray}
a_i^{\dagger}a_k^{\dagger}|0\rangle&=&|\ldots 1_i\ldots 1_k\rangle%
\nonumber\\%
a_k^{\dagger}a_i^{\dagger}|0\rangle&=&-|\ldots 1_i\ldots 1_k\rangle%
\label{2.14}
\end{eqnarray}
The trick is a consequence of the different order in disclosing the Slater determinant. In the first line, we expand the determinant from the main diagonal, but in the second line the columns are permuted, and the determinant changes sign.

Returning to our guideline, we define the following order in the set of the spin state, which we are interested in,
$$1H<1\bar{H}<2H<2\bar{H}$$ 
We now reconsider the above transformations for bosonic quasiparticles.
Instead of (\ref{2.5}) we obtain
\begin{eqnarray}
|\psi\rangle_{\mathrm{out}}&=&\hat{S}a^{\dagger}_{1X}a^{\dagger}_{2X}|0\rangle
=\frac{1}{2}\left(b^{\dagger}_{1X}+b^{\dagger}_{2X}\right)\left(b^{\dagger}_{1X}-b^{\dagger}_{2X}\right)|0\rangle%
\nonumber\\%
&=&\frac{1}{2}\left[-b^{\dagger}_{1X}b^{\dagger}_{2X}+b^{\dagger}_{2X}b^{\dagger}_{1X}\right]|0\rangle=-b^{\dagger}_{1X}b^{\dagger}_{2X}|0\rangle%
\nonumber\\%
\nonumber\\%
&=&-|1_{1X},1_{2X}\rangle%
\label{2.15}
\end{eqnarray}
or explicitly
\begin{eqnarray}
|1_{1H},1_{2H}\rangle&\to&-|1_{1H},1_{2H}\rangle%
\nonumber\\%
|1_{1\bar{H}},1_{2\bar{H}}\rangle&\to&-|1_{1\bar{H}},1_{2\bar{H}}\rangle%
\label{2.16}%
\end{eqnarray}
Both states, being triplet states of the total spin angular momentum with projections $\pm 1$, are eigenstates for the fermionic beamsplitter.

We can define a triplet spin state with zero spin projection for two fermions
\begin{equation}
|\psi\rangle_{\mathrm{in}}=\frac{1}{\sqrt{2}}\left[|1_{1H},1_{2\bar{H}}^{\phantom{\dagger}}\rangle+|1_{1\bar{H}},1_{2H}\rangle\right]%
\label{2.17}
\end{equation}
This state is constructed similarly to (\ref{2.7}), but unlike (\ref{2.8}), we expect it to be conserved
\begin{eqnarray}
|\psi\rangle_{\mathrm{out}}&=&\hat{S}\frac{1}{\sqrt{2}}\left[a^{\dagger}_{1H}a^{\dagger}_{2\bar{H}}+a^{\dagger}_{1\bar{H}}a^{\dagger}_{2H}\right]|0\rangle%
\nonumber\\%
&=&\frac{1}{2\sqrt{2}}\left[\left(b^{\dagger}_{1H}+b^{\dagger}_{2H}\right)\left(b^{\dagger}_{1\bar{H}}-b^{\dagger}_{2\bar{H}}\right)\right.%
\nonumber\\%
&&\left.+\left(b^{\dagger}_{1\bar{H}}+b^{\dagger}_{2\bar{H}}\right)\left(b^{\dagger}_{1H}-b^{\dagger}_{2H}\right)\right]|0\rangle%
\nonumber\\%
&=&\frac{1}{\sqrt{2}}\left[-b^{\dagger}_{1H}b^{\dagger}_{2\bar{H}}-b^{\dagger}_{1\bar{H}}b^{\dagger}_{2H}\right]|0\rangle%
\nonumber\\%
&=&-\frac{1}{\sqrt{2}}\left[|1_{1H},1_{2\bar{H}}^{\phantom{\dagger}}\rangle+|1_{1\bar{H}},1_{2H}\rangle\right]%
\label{2.18}
\end{eqnarray}
and $|\psi\rangle_{\mathrm{out}}=-|\psi\rangle_{\mathrm{in}}$ i.e. this state is indeed an eigenstate of the beamsplitter.

Finally let us follow what happens with the singlet state
\begin{equation}
|\psi\rangle_{\mathrm{in}}=\frac{1}{\sqrt{2}}\left[|1_{1H},1_{2\bar{H}}^{\phantom{\dagger}}\rangle-|1_{1\bar{H}},1_{2H}\rangle\right]%
\label{2.19}
\end{equation}
Then we obtain
\begin{eqnarray}
|\psi\rangle_{\mathrm{out}}&=&\hat{S}\frac{1}{\sqrt{2}}\left[a^{\dagger}_{1H}a^{\dagger}_{2\bar{H}}-a^{\dagger}_{1\bar{H}}a^{\dagger}_{2H}\right]|0\rangle%
\nonumber\\%
&=&\frac{1}{2\sqrt{2}}\left[\left(b^{\dagger}_{1H}+b^{\dagger}_{2H}\right)\left(b^{\dagger}_{1\bar{H}}-b^{\dagger}_{2\bar{H}}\right)\right.%
\nonumber\\%
&&\left.-\left(b^{\dagger}_{1\bar{H}}+b^{\dagger}_{2\bar{H}}\right)\left(b^{\dagger}_{1H}-b^{\dagger}_{2H}\right)\right]|0\rangle%
\nonumber\\%
&=&\frac{1}{\sqrt{2}}\left[b^{\dagger}_{1H}b^{\dagger}_{1\bar{H}}-b^{\dagger}_{2H}b^{\dagger}_{2\bar{H}}\right]|0\rangle%
\nonumber\\%
&=&+\frac{1}{\sqrt{2}}\left[|1_{1H},1_{1\bar{H}}^{\phantom{\dagger}}\rangle-|1_{2H},1_{2\bar{H}}\rangle\right]%
\label{2.20}
\end{eqnarray}
or explicitly
\begin{eqnarray}
\lefteqn{\frac{1}{\sqrt{2}}\left[|1_{1H},1_{2\bar{H}}^{\phantom{\dagger}}\rangle-|1_{1\bar{H}},1_{2H}\rangle\right]}%
\nonumber\\%
&\to&\frac{1}{\sqrt{2}}\left[|1_{1H},1_{1\bar{H}}^{\phantom{\dagger}}\rangle-|1_{2H},1_{2\bar{H}}\rangle\right]%
\label{2.21}%
\end{eqnarray}
which reveals transformation to a singlet state but assisted by bunching of the particles in the output channels.

\subsubsection{Clarifying remarks}

\noindent 
To summarize this part, we can point out the following:(i) There is an evident physical similarity in behavior of non-interacting  bosons and fermions under beamsplitter-type transformation. (ii) The symmetric polarization state of bosons becomes bunched after the interaction, but the antisymmetric state remains unchanged. (iii) And vice versa, unlike the bosons, the fermions save their triplet states, but have bunched the singlet state. 

This quantum conjugation between spin and positions of identical particles is a fundamental property which manifests itself in various physical processes. In the present report, we aim to focus on the phenomena of electron bunching and antibunching as a key ingredient in construction of covalent bonds in chemistry. The qualitative analysis of chemical forces uses a molecular orbital approach to imagine binding interactions in a molecule. In such approach the electrons, considered as independent particles, driven by a self-consistent effective potential, are placed in the molecular orbitals, arrayed from the lower to excited energies, with keeping in mind the Pauli exclusion principle and Hund's rule of maximum multiplicity (only two electrons, having opposite spins, per orbital; place as many unpaired electrons on one energy level as possible before starting to pair them). 

We foresee a clear similarity with the position bunching of fermions, observed by Hong-Ou-Mandel interference for the singlet state. And in this regard, the above overview, towards comparison of the Hong-Ou-Mandel interference for different quantum statistics, highlights a basic equivalence in visualization of spin vs. position quantum conjugation for either statistic at least for a system consisting of two particles. 

\begin{figure}[tp]
\includegraphics[width=8 cm]{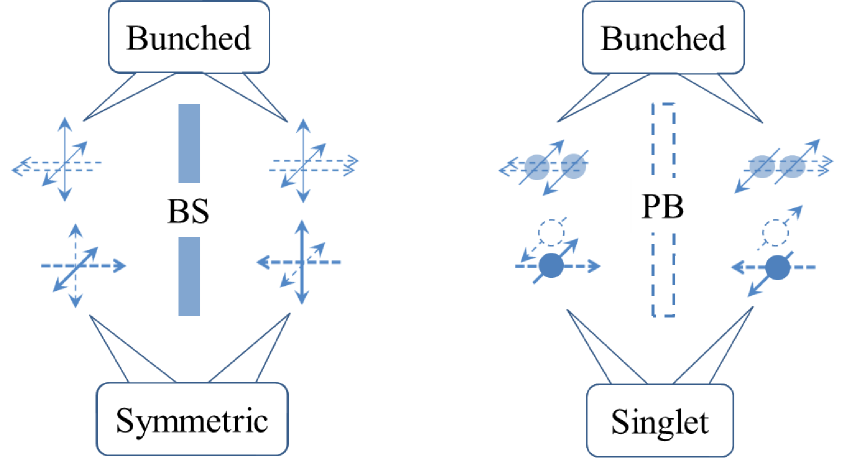}
\caption{Diagram visualization of transformations (\ref{2.9}) (left) and (\ref{2.21}) (right), both leading to bunching of either photons, prepared in a symmetric polarization-entangled state and mixed by a beamsplitter (BS), or electrons, incoming in a singlet spin-entangled state and scattered by potential barrier (PB), see text. The inward single arrows and outward double arrows indicate the wave-vectors, for either photons or electrons, respectively before and after scattering. The orthogonal arrows indicate their vector and spin polarizations.}
\label{fig1}%
\end{figure}

As a supporting example, in Fig.~\ref{fig1}(left) transformation (\ref{2.9}) is shown as a beamsplitter (BS) mixing of two photons originated in a symmetric polarization-entangled state. In Fig.~\ref{fig1}(right) it is compared with one-dimensional scattering of two electrons, prepared and incoming in a singlet spin-entangled  state, on a potential barrier (PB). The output states are physically identical for both the processes and reveal bunched states for outgoing photons and electrons.

As we further see, certain combinations of several bonding molecular orbitals, occupied by electrons, in some cases can be fairly simulated by a system of neutral atoms, prepared in special quantum states. The specifics of bunching or antibunching in atomic systems, outlined in the next section, is that the atoms are typically bosons while their pseudospin subsystem, constructed as an entangled array of coupled one-half pseudospins, is physically equivalent to a spin subsystem of coupled fermions. We are aiming to give non-trivial examples where the collections consisting of either electrons or atoms, existing in the same environment, would have equivalent position distributions conjugated with their spin states.

\section{Manipulation with the trapped bosonic atoms }\label{Section_III}

\noindent Here, we refer to an ideal but still experimentally attainable configuration. We assume a system of two atoms slowed down to the ground vibrational mode of the capturing tweezers. The atomic spin states can be entangled by the protocol of Rydberg blockade to an arbitrary either Bell-type or triplet and singlet states. We simplify the problem by considering atoms initially loaded into two identical microtraps and neglecting the interaction between them.

\subsection{Description of the model}

\begin{figure}[tp]
\includegraphics[width=8 cm]{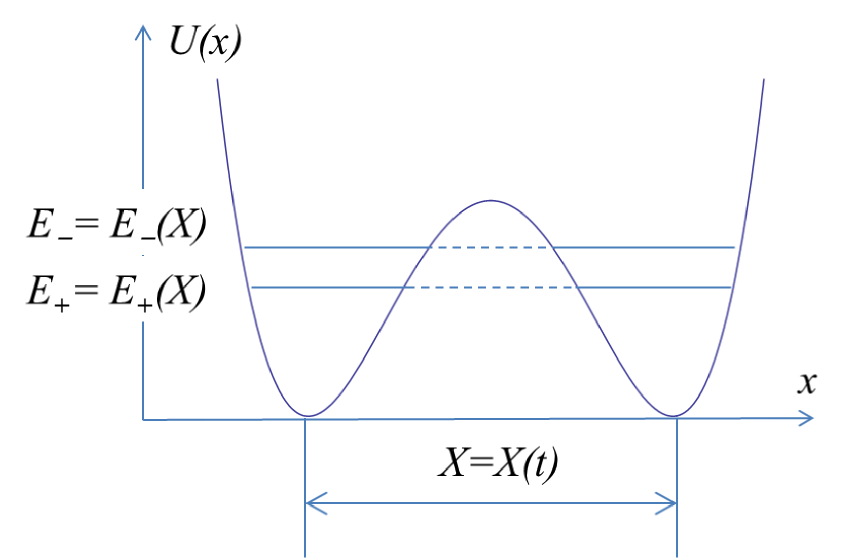}
\caption{One-dimensional visualization of a double well microtrap potential, see text.}
\label{fig2}%
\end{figure}

\noindent  
Consider two atoms, each held in an individual trap, with potential wells that can overlap and interfere.
The traps are separated by a distance $X$, then the potential shape and energy structure are shown in Fig.~\ref{fig2} in one-dimensional visualization.  
If the original tweezers are identical, then each pair of energy states existing in the individual wells splits into two states, $E_{+}$ and $E_{-}$.
The splitting depends on the separation $X=X(t)$, which is adiabatically slow varied in time, and the state with lower (upper) energy belongs to gerade (ungerade) symmetry. 

In the second quantized formalism, with neglecting the direct interaction between the atoms and for a fixed separation $X$, the system can be described by the following Hamiltonian
\begin{equation}
\hat{H}=\int dx\left[-\frac{\hbar^2}{2m}\psi^{\dagger}(x)\triangle\psi(x)+\psi^{\dagger}(x)U(x)\psi(x)\right]%
\label{3.1}
\end{equation}
where the reference operators are given by
\begin{eqnarray}
\psi(x)&=&\sum_{\alpha=a,b}\phi_{+}(x)\chi_\alpha\,a_{+\alpha}+\phi_{-}(x)\chi_\alpha\,a_{-\alpha}%
\nonumber\\%
\psi^{\dagger}(x)&=&\sum_{\alpha=a,b}\phi^{\ast}_{+}(x)\chi^{\dagger}_\alpha\,a^{\dagger}_{+\alpha}+\phi^{\ast}_{-}(x)\chi^{\dagger}_\alpha\,a^{\dagger}_{-\alpha}%
\label{3.2}%
\end{eqnarray}
and wavefunctions  $\phi_{+}(x)$ and $\phi_{-}(x)$ are given by solution of the stationary Schr\"{o}dinger equation
\begin{equation}
-\frac{\hbar^2}{2m}\triangle\,\phi_{\pm}(x)+U(x)\phi_{\pm}(x)=E_{\pm}\phi_{\pm}(x)%
\label{3.3}
\end{equation}
and parameterized by the tweezers separation $X$, such that $E_{\pm}=E_{\pm}(X)$. The pseudospin states $\chi_{\alpha}\equiv|\alpha\rangle$ suggest the clock states with $|\alpha\rangle=|a\rangle$ (spin down) and $|\alpha\rangle=|b\rangle$ (spin up).\footnote{Here we have in mind the alkali-metal atoms, normally used in a quantum simulator, such that $|a\rangle$=$|F_{-},0\rangle$ and $|b\rangle$=$|F_{+},0\rangle$, which are the Zeeman states respectively belonging to the lower and upper hyperfine sublevels of the ground state, and having zero projection of the total spin angular momentum.}

Substitute (\ref{3.2}) into (\ref{3.1}) and transfer the Hamiltonian to another form
\begin{equation}
\hat{H}(X)=\sum_{\alpha}\left[E_{+}(X)\,a^{\dagger}_{+\alpha}a_{+\alpha}+E_{-}(X)\,a^{\dagger}_{-\alpha}a_{-\alpha}\right]%
\label{3.4}
\end{equation}
where we have used the normalization $\chi^{\dagger}_\alpha\chi_\alpha=1$ and highlighted the dependence on separation $X$ as on external parameter.

\subsection{The Heisenberg dynamics for an adiabatically drifted Hamiltonian }\label{Section_IIIB}

\noindent Let the separation between the trap wells depend on time $X=X(t)$. Then, in conventional description, the dependence on time is incorporated into a chronologically $T$-ordered evolutionary operator $\hat{\cal S}(t,0)$ and expressed by transformation of the system wavefunction
\begin{eqnarray}
|\psi(t)\rangle&=&\hat{\cal S}(t,0)|\psi(0)\rangle
\nonumber\\%
\hat{\cal S}(t,0)&=&T\exp\left[-\frac{i}{\hbar}\int_0^t\hat{H}(t')dt'\right]
\label{3.5}%
\end{eqnarray}
Then in Eq.~(\ref{3.1}) $U(x)\to U(x;X(t))\equiv U(x,t)$, but in Eqs.~(\ref{3.2}),(\ref{3.3}) we have to accept a particular basis functions, for example, associated with the potential profile taken at initial time. As a consequence, the transformation of the system Hamiltonian to (\ref{3.4}) is no longer fulfilled. The initial state vector $|\psi(0)\rangle$ is defined for a specific quantum configuration at a given moment in time.

However, if the parameter $X=X(t)$ is a slow varying external quantity, the inconvenience of such a conventional approach is evident. If, for a while, we forget about the time dynamics, we can keep the equations (\ref{3.1})--(\ref{3.4}) and parameterize them by $X=X(t)$. The subtle point is how to define the reference operators, namely $a_{\pm\alpha}$ and $a^{\dagger}_{\pm\alpha}$. Without loss of generality we can fix these operators by the following single particle matrix elements
\begin{eqnarray}
\langle 0|a_{\pm\alpha}|\ldots,1_{\pm\alpha}(t),\ldots\rangle&=&1%
\nonumber\\%
\langle\ldots,1_{\pm\alpha}(t),\ldots|a^{\dagger}_{\pm\alpha}|0\rangle&=&1%
\label{3.6}%
\end{eqnarray}
where the time argument indicates that the single particle state is configured as an eigenstate of the Hamiltonian at a given time $t$:
\begin{equation}
\hat{H}\left[X(t)\right]\,|1_{\pm\alpha}(t)\rangle=E_{\pm}\left[X(t)\right]\,|1_{\pm\alpha}(t)\rangle%
\label{3.7}%
\end{equation}
It is important that the defined operators $a_{\pm\alpha}$ and $a^{\dagger}_{\pm\alpha}$ stay independent on time. They only reflect the fact of creation and annihilation events in a many particle configuration of non-interacting atoms evolved up to arbitrary time $t$. So the Hamiltonian contributed in (\ref{3.7}) is saved in form (\ref{3.4})
\begin{equation}
\hat{H}\left[X(t)\right]=\sum_{\alpha}\left[E_{+}\left[X(t)\right]\,a^{\dagger}_{+\alpha}a_{+\alpha}+E_{-}\left[X(t)\right]\,a^{\dagger}_{-\alpha}a_{-\alpha}\right]%
\label{3.8}
\end{equation}
but parameterized by time.

Now consider the time dynamics for wavefunction (\ref{3.5}) driven by the Schrodinger equation
\begin{equation}
\mathrm{i}\hbar\,\frac{d}{dt}|\psi(t)\rangle=\hat{H}\left[X(t)\right]\,|\psi(t)\rangle%
\label{3.9}%
\end{equation}
Again without loss of generality we can simplify it by a single particle case with $|\psi(0)\rangle=|1_{\pm\alpha}(0)\rangle\equiv|1_{\pm\alpha}\rangle$. Instead of eigenstate at initial time $|1_{\pm\alpha}\rangle$, let us express the evolved state vector $|\psi(t)\rangle$ by the eigenstate $|1_{\pm\alpha}(t)\rangle$ at a given time. Then we obtain
\begin{equation}
|\psi(t)\rangle=\exp\left\{-\frac{i}{\hbar}\int_0^t\! E_{\pm}\left[X(t')\right]dt'+i\gamma_{\pm}(t)\right\}\,|1_{\pm\alpha}(t)\rangle%
\label{3.10}%
\end{equation}
where
\begin{equation}
\gamma_{\pm}(t)=i\int_0^t dt\,\langle 1_{\pm\alpha}(t)|\nabla_{X}1_{\pm\alpha}(t)\rangle\,\dot{X}(t)%
\label{3.11}%
\end{equation}
is so called geometrical phase. Here we have taken into account that the eigenstate depends on time only via the separation $X=X(t)$ and assumed that, in general, it can be a vector quantity.

In accordance with (\ref{3.6}) and (\ref{3.10}) we arrive at
\begin{eqnarray}
\langle 0|a_{\pm\alpha}|\psi(t)\rangle&=&\exp\left\{-\frac{i}{\hbar}\int_0^t\! E_{\pm}\left[X(t')\right]dt'+i\gamma_{\pm}(t)\right\}%
\nonumber\\%
\langle\psi(t)|a^{\dagger}_{\pm\alpha}|0\rangle&=&\exp\left\{+\frac{i}{\hbar}\int_0^t\! E_{\pm}\left[X(t')\right]dt'-i\gamma_{\pm}(t)\right\}%
\nonumber\\%
\label{3.12}%
\end{eqnarray}
These matrix elements can be alternatively written as
\begin{eqnarray}
\langle 0|a_{\pm\alpha}|\psi(t)\rangle&=&\langle 0|a_{\pm\alpha}(t)|1_{\pm\alpha}(0)\rangle%
\nonumber\\%
\langle\psi(t)|a^{\dagger}_{\pm\alpha}|0\rangle&=&\langle 1_{\pm\alpha}(0)|a^{\dagger}_{\pm\alpha}(t)|0\rangle%
\label{3.13}%
\end{eqnarray}
where we have transferred the adiabatically induced dynamics onto the following Heisenberg-type operators
\begin{eqnarray}
a_{\pm\alpha}(t)&=&\exp\left\{-\frac{i}{\hbar}\int_0^t\! E_{\pm}\left[X(t')\right]dt'+i\gamma_{\pm}(t)\right\}a_{\pm\alpha}%
\nonumber\\%
a^{\dagger}_{\pm\alpha}(t)&=&\exp\left\{+\frac{i}{\hbar}\int_0^t\! E_{\pm}\left[X(t')\right]dt'-i\gamma_{\pm}(t)\right\}a^{\dagger}_{\pm\alpha}%
\nonumber\\%
\label{3.14}%
\end{eqnarray}
and referred the basis states with conventionally chosen "zero-time". We will further use the defined operators to construct a transformation for an artificially controllable "collision" of two atoms, each originally confined and guided by its individual microtrap.

\subsection{Hong-Ou-Mandel-type transformations with overlapped atoms}\label{Section_IIIC}

\noindent Let us introduce two asymptotes for these operators. For infinite past with $t\to -\infty$ we define
\begin{eqnarray}
\exp\left[+\frac{i}{\hbar}E_0t\right]a_{+\alpha}(t)&\equiv&a^{(\mathrm{in})}_{+\alpha}=\frac{1}{\sqrt{2}}\left[a_{1\alpha}+a_{2\alpha}\right]%
\nonumber\\%
\exp\left[+\frac{i}{\hbar}E_0t\right]a_{-\alpha}(t)&\equiv&a^{(\mathrm{in})}_{-\alpha}=\frac{1}{\sqrt{2}}\left[a_{1\alpha}-a_{2\alpha}\right]%
\nonumber\\%
\label{3.15}%
\end{eqnarray}
where $E_0=\lim_{X\to\infty}E_{\pm}(X)$ is the state energy of independent atoms.  In accordance with (\ref{3.14}), the operators $a^{(\mathrm{in})}_{\pm\alpha}$ are independent of time and insensitive to the separation distance at the infinite limit $X(t)\to\infty$. The last equalities in (\ref{3.15}) expand these "in"-asymptotes in terms of the Schr\"{o}dinger operators of annihilation of an atom at each site either $1$ or $2$. The Hermitian conjugated transformations can be performed for the creation operators.

For infinite future with $t\to +\infty$ we define
\begin{eqnarray}
\exp\left[+\frac{i}{\hbar}E_0t\right]a_{+\alpha}(t)&\equiv&a^{(\mathrm{out})}_{+\alpha}=\frac{1}{\sqrt{2}}\left[a^{(\mathrm{out})}_{1\alpha}+a^{(\mathrm{out})}_{2\alpha}\right]%
\nonumber\\%
\exp\left[+\frac{i}{\hbar}E_0t\right]a_{-\alpha}(t)&\equiv&a^{(\mathrm{out})}_{-\alpha}=\frac{1}{\sqrt{2}}\left[a^{(\mathrm{out})}_{1\alpha}-a^{(\mathrm{out})}_{2\alpha}\right]%
\nonumber\\%
\label{3.16}%
\end{eqnarray}
which introduces the operators of "out"-asymptotes. 
The trick is that, although formally treated as site operators $a^{(\mathrm{out})}_{1\alpha}$ and $a^{(\mathrm{out})}_{2\alpha}$,  and being time-independent Schr\"{o}dinger-type operators, they have accumulated the transformation following the artificial collision process, whose crucial step involved a mandatory overlap of the potential wells.

The site indices $1$ and $2$ are only conventionally used in (\ref{3.16}) and should not be confused here with the real occupation of the respective site. Due to intrinsically dynamical nature of such an artificially constructed collision, these operators could be linked with the "in"-operators by a certain unitary transformation
\begin{eqnarray}
a^{(\mathrm{out})}_{1\alpha}&=& \hat{\mathfrak{S}}\,a_{1\alpha}\,\hat{\mathfrak{S}}^{\dagger}%
\nonumber\\%
a^{(\mathrm{out})}_{2\alpha}&=& \hat{\mathfrak{S}}\,a_{2\alpha}\,\hat{\mathfrak{S}}^{\dagger}%
\label{3.17}%
\end{eqnarray}
where $\mathfrak{S}$-operator plays the same role for the artificial collision as the scattering $S$-matrix for real collision.

From (\ref{3.14}) we obtain
\begin{eqnarray}
a^{(\mathrm{out})}_{+\alpha}&=&\mathrm{e}^{i\theta_{+}}\,a^{(\mathrm{in})}_{+\alpha}%
\nonumber\\%
a^{(\mathrm{out})}_{-\alpha}&=&\mathrm{e}^{i\theta_{-}}\,a^{(\mathrm{in})}_{-\alpha}%
\label{3.18}%
\end{eqnarray}
where
\begin{eqnarray}
\theta_{+}&=&-\int_{-\infty}^{+\infty}dt\,\Delta_{+}\left[X(t)\right] + \gamma_{+}(C)%
\nonumber\\%
\theta_{-}&=&-\int_{-\infty}^{+\infty}dt\,\Delta_{-}\left[X(t)\right] + \gamma_{-}(C)%
\label{3.19}
\end{eqnarray}
where
\begin{eqnarray}
\Delta_{+}\left[X(t)\right]&=&\frac{1}{\hbar}\left\{E_{+}\left[X(t)\right]-E_0\right\}%
\nonumber\\%
\Delta_{-}\left[X(t)\right]&=&\frac{1}{\hbar}\left\{E_{-}\left[X(t)\right]-E_0\right\}%
\nonumber\\%
\label{3.20}%
\end{eqnarray}
and $\Delta_{+}\left[X(t)\right]\approx-\Delta_{-}\left[X(t)\right]$ with high accuracy. However, in general $\theta_{+}\neq -\theta_{-}$ just due to the geometrical phase.

The latter appears when the contours $C$ shaped by the converging and diverging paths of the overall process are different and $ \gamma_{+}(C)\sim \gamma_{-}(C)$. If we assume the same incoming and outgoing paths then the contoured area has zero measure and the geometric phase vanishes. In this case, which we will further assume, $\theta_{+}= -\theta_{-}\equiv\theta$.

With making use of (\ref{3.18}) and (\ref{3.17}) we arrive at the following "in"-to-"out" transformation for the in-situ creation operators
\begin{eqnarray}
\hat{\mathfrak{S}}\,a_{1\alpha}^{\dagger}\,\hat{\mathfrak{S}}^{\dagger}&=&\cos\theta\,a_{1\alpha}^{\dagger}-\mathrm{i}\sin\theta\,a_{2\alpha}^{\dagger}%
\nonumber\\%
\hat{\mathfrak{S}}\,a_{2\alpha}^{\dagger}\,\hat{\mathfrak{S}}^{\dagger}&=&-\mathrm{i}\sin\theta\,a_{1\alpha}^{\dagger}+\cos\theta\,a_{2\alpha}^{\dagger}%
\label{3.21}%
\end{eqnarray}
which is unitary transformation identical to those we have already considered in the preceding Section \ref{Section_II}.\footnote{The contribution of the  original Schr\"{o}dinger operators in the right-hand side of (\ref{3.21}), unlike to (\ref{2.4}), is a direct consequence of physical equivalence between input and output scattering channels for this case.}

\subsubsection{Triplet states}

\noindent Take $\theta=\pi/4$ in (\ref{3.21}). Then instead of (\ref{2.5}) we obtain
\begin{eqnarray}
|\psi\rangle_{\mathrm{out}}&=&\hat{\mathfrak{S}}a^{\dagger}_{1\alpha}a^{\dagger}_{2\alpha}|0\rangle
=\frac{1}{2}\left(a^{\dagger}_{1\alpha}-\mathrm{i}\,a^{\dagger}_{2\alpha}\right)\left(-\mathrm{i}\,a^{\dagger}_{1\alpha}+a^{\dagger}_{2\alpha}\right)|0\rangle%
\nonumber\\%
&=&-\frac{\mathrm{i}}{2}\left[a^{\dagger}_{1\alpha}a^{\dagger}_{1\alpha}+a^{\dagger}_{2\alpha}a^{\dagger}_{2\alpha}\right]|0\rangle%
=-\frac{\mathrm{i}}{\sqrt{2}}\left[|2_{1\alpha}^{\phantom{\dagger}}\rangle+|2_{2\alpha}\rangle\right]%
\nonumber\\%
\label{3.22}
\end{eqnarray}
and instead of (\ref{2.8})
\begin{eqnarray}
|\psi\rangle_{\mathrm{out}}&=&\hat{\mathfrak{S}}\frac{1}{\sqrt{2}}\left[a^{\dagger}_{1a}a^{\dagger}_{2b}+a^{\dagger}_{1b}a^{\dagger}_{2a}\right]|0\rangle%
\nonumber\\%
&=&\frac{1}{2\sqrt{2}}\left[\left(a^{\dagger}_{1a}-\mathrm{i}\,a^{\dagger}_{2a}\right)\left(-\mathrm{i}\,a^{\dagger}_{1b}+a^{\dagger}_{2b}\right)\right.%
\nonumber\\%
&&\left.+\left(a^{\dagger}_{1b}-\mathrm{i}\,a^{\dagger}_{2b}\right)\left(-\mathrm{i}\,a^{\dagger}_{1a}+a^{\dagger}_{2a}\right)\right]|0\rangle%
\nonumber\\%
&=&-\frac{\mathrm{i}}{\sqrt{2}}\left[a^{\dagger}_{1a}a^{\dagger}_{1b}+a^{\dagger}_{2a}a^{\dagger}_{2b}\right]|0\rangle%
\nonumber\\%
&=&-\frac{\mathrm{i}}{\sqrt{2}}\left[|1_{1a},1_{1b}^{\phantom{\dagger}}\rangle+|1_{2a},1_{2b}\rangle\right]%
\label{3.23}
\end{eqnarray}
Both the cases have a clear signature of the Hong-Ou-Mandel bunching effect.

\subsubsection{Singlet state}

\noindent Instead of (\ref{2.11}) we obtain
\begin{eqnarray}
|\psi\rangle_{\mathrm{out}}&=&\hat{\mathfrak{S}}\frac{1}{\sqrt{2}}\left[a^{\dagger}_{1a}a^{\dagger}_{2b}-a^{\dagger}_{1b}a^{\dagger}_{2a}\right]|0\rangle%
\nonumber\\%
&=&\frac{1}{2\sqrt{2}}\left[\left(a^{\dagger}_{1a}-\mathrm{i}\,a^{\dagger}_{2a}\right)\left(-\mathrm{i}\,a^{\dagger}_{1b}+a^{\dagger}_{2b}\right)\right.%
\nonumber\\%
&&\left.-\left(a^{\dagger}_{1b}-\mathrm{i}\,a^{\dagger}_{2b}\right)\left(-\mathrm{i}\,a^{\dagger}_{1a}+a^{\dagger}_{2a}\right)\right]|0\rangle%
\nonumber\\%
&=&\frac{1}{\sqrt{2}}\left[a^{\dagger}_{1a}a^{\dagger}_{2b}-a^{\dagger}_{1b}a^{\dagger}_{2a}\right]|0\rangle%
\nonumber\\%
&=&\frac{1}{\sqrt{2}}\left[|1_{1a},1_{2b}^{\phantom{\dagger}}\rangle-|1_{1b},1_{2a}\rangle\right]%
\label{3.24}
\end{eqnarray}
The state is unchanged, being an antibunched eigenstate of the transformation.
 
With neglected interatomic interaction, all three states (\ref{3.22})--(\ref{3.24}) superpose the partial contributions having the same energies and for ideal scenarios: (i) The protocol can be adjusted even for a single atom. Then it prepares a spatial qubit where the logic "0" and "1" are associated with the atom's location; (ii) The states (\ref{3.23}) and (\ref{3.24}) are entangled, but the state (\ref{3.22}) is disentangled with the spin subsystem and easier for preparation; (iii) The state (\ref{3.22}) can be suggested as a $NOON$-type qubit state constructed with two atoms. The key advantage is that here there is an extra resource -- two matter pieces in qubit arms -- useful for its further processing under error correction protocols.

\subsubsection{Relation with fermions and covalent bonding}

\noindent As we see, in the system of two atoms, their bosonic nature does not prevent a realistic simulation of any mutual position distribution of two fermions existing in the same environment. One only needs to properly select the pseudospin state and trace the density matrix over the pseudospin variables. In particular, in the middle part of the overlapping process, the bunched states (\ref{3.22}) and (\ref{3.23}) adiabatically correlate with the occupied orbitals having enhanced matter density between the trapping centers (see Fig.~\ref{fig2}), i.e. relevantly reproduce a covalent bond. Unlike electrons, where the density enhancement is provided by a singlet spin state, see (\ref{a.1})-(\ref{a.3}), for atoms, for the bunched states, there is no need in pseudospin entanglement for experimental verification of adiabatic transformation (\ref{3.21}).

It might seem that it would be the only relevant example of two atoms where such a parity in description of fermions and bosons, mediated by their collective spin or pseudospin states, would be possible. 
Moreover, once we added the neglected interaction between the atoms, we would partly disturb the above ideal scenario. Nevertheless, as commonly known, the non-relativistic description of electronic configuration in a molecule is critically based on a quantum conjugation between the collective spin state of the electrons and their position variables. That gives us a chance for searching for the physical similarity in position behavior of fermions and bosons in more complex systems at least for specific values of the total spin angular momentum. 
 
As we further show, the compound and stable fermionic configurations consisting of three and four electrons, with minimal total spin angular momentum, can be faithfully reproduced by bosonic atoms.
That justifies the use of atoms to emulate electrons in an analogue quantum simulation of their internal dynamics toward the formation of divalent bonds in organic molecules.

\section{The matter wave interference of three and four particles}\label{Section_IV}

\subsection{The spin controllable matter wave interference}

\noindent 
The electronic shells of large molecules typically have a closed spin subsystem with a minimal total spin
\cite{ZahradníkPolak1980}. 
As clarified in the Appendix \ref{Appendix_A}, 
if a degenerate system consists of three or four particles, there exists a unique configuration for either spin or pseudospin states, conjugated with the spatial variables, such that, under the same external conditions, the position behavior of fermions and bosons is described similarly.
Then the quantum dynamics of an electron enclave, developing on atomic scale, can be realistically replicated by a much larger mesoscopically scaled quantum object consisting of bosonic atoms. 

Of course, for a realistic scenario, it would make a difference how constructively the environment action and internal interaction were reproduced by the atomic system. Strictly speaking, we suggest not ideal quantum simulations, but instead a certain "Quantum Lego" protocol with the option of only qualitative reproduction of important physical details. The basic concept of the protocol is to prepare an initial entangled state of separated atomic pseudospins and then to track the state evolution by spatial transfer of the microtraps transporting the atoms. After a round of internal interactions, the traps can be separated for remote and independent measurements. 
The modified final state and the results of measurements would provide information about the processes that occurred during the interaction stage.

As an example, for the system of three atoms, originally occupying three sites $A$, $B$, $C$, we can suggest the following probe state with total pseudospin $1/2$, constructed by adding a third particle to a pair coupled to $0$ pseudospin
\begin{widetext}
\begin{eqnarray}
\Psi^{(0;1/2)}(1,2,3)&=&\frac{1}{2\sqrt{6}}\alpha(1)[\beta(2)\gamma(3)-\beta(3)\gamma(2)]\varphi_A(1)[\varphi_B(2)\varphi_C(3)-\varphi_B(3)\varphi_C(2)]%
\nonumber\\%
&+&\frac{1}{2\sqrt{6}}\alpha(2)[\beta(3)\gamma(1)-\beta(1)\gamma(3)]\varphi_A(2)[\varphi_B(3)\varphi_C(1)-\varphi_B(1)\varphi_C(3)]%
\nonumber\\%
&+&\frac{1}{2\sqrt{6}}\alpha(3)[\beta(1)\gamma(2)-\beta(2)\gamma(1)]\varphi_A(3)[\varphi_B(1)\varphi_C(2)-\varphi_B(2)\varphi_C(1)]%
\label{4.1}%
\end{eqnarray}
\end{widetext}
where $\alpha(i)$, $\beta(j)$, $\gamma(k)$ denote the basis spin functions and $\varphi_A(i)$, $\varphi_B(j)$, $\varphi_C(k)$ denote the original in-situ orbitals. Here, $i,j,k$ denote both the spin and position arguments running $1,2,3$ and subsequently changed by the permutation rules. This state defines a fully symmetric representation for three atoms, but contains the singlet-type pseudospin entanglement prepared on sites $B$ and $C$. 
This entanglement can be provided by a Rydberg blockade protocol. Note that there are various options for entangling other site pairs. In a similar way, we can construct probe singlet states in the system of four atoms. 

Further dynamics of such probe states reveals a complex quantum process, which would be not so easy to follow for classical computational algorithms, but it could be naturally followed by quantum simulations. Indeed, the key feature of time dynamics in the non-relativistic regime is that the atomic wavefuncion $\Psi^{(b)}=\Psi^{(b)}(\ldots;t)$, defined by (\ref{a.26}), would preserve its superposition structure but would take dependence on time in the expansion coefficients $C_1=C_1(t)$ and $C_2=C_2(t)$ as well as in the basis functions $\Psi_{1}^{(b)}(\ldots;t)$ and $\Psi_{2}^{(b)}(\ldots;t)$. The latter results from the internal adiabatic transformations under the site transfers. 

The dynamics, being a natural attribute of the quantum transformation, should be highly sensitive to the tracked paths of the sites and the speed of the transfer. For example, if for the initial state (\ref{4.1}) one intends to make closer only sites $B$ and $C$ then the suggested state would be an eigenstate of the process, and after separation we would arrive at the same state, see the preceding section. 
However, if all three sites are overlapped, the protocol would lead to spin entanglement of all of them. In general, any initial state should be re-expanded in the basis of the eigenstates for a particular trap configuration created by the transferring paths. Each such path, driving the system evolution, would be described by its unique unitary transformation in the entire Hilbert subspace. 

The proposed quantum simulator assumes that the joint charge distribution of electrons, described by the fermion wavefunction $\Psi^{(f)}=\Psi^{(f)}(\ldots;t)$, given by (\ref{a.25}), under similar conditions would behave similarly if we associate it with alternative conjugation of the spin and position variables, see Appendices \ref{Appendix_A} and \ref{Appendix_B}. The main chemical creature of dynamical evolution is in either construction or destruction of binding bonds, which is ruled by time behavior of the joint probability distribution and by the second order quantum interference, associated with either bunching or anti-bunching effects. Below we present a round of illustrative calculations for the position wavefunctions estimated in the orbital model, detailed in the Appendix \ref{Appendix_B}.

\subsection{Joint probability distributions in MO-LCAO approximation}

\begin{figure}[tp]
\includegraphics[width=5 cm]{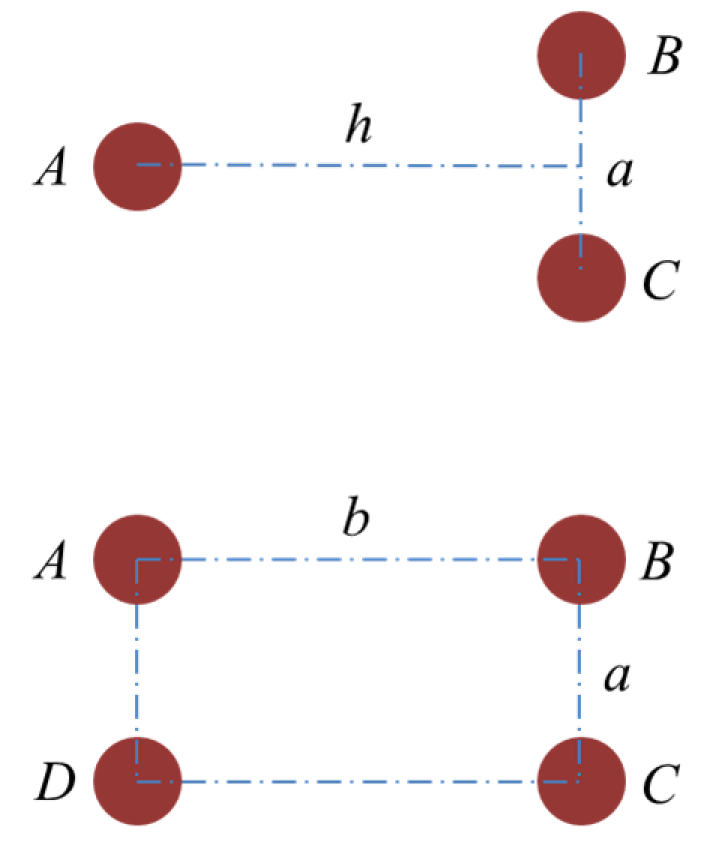}
\caption{The considered site configurations for systems of three and four particles, see text.}
\label{fig3}%
\end{figure}

\noindent Let us focus on the energy ground state and restrict ourselves by two symmetric configurations, namely, by isosceles triangle (three atoms) and by rectangular (four atoms), as shown in Fig.~\ref{fig3}. The probe eigenfunctions can be taken in a Roothaan basis set of the single particle orbital functions \cite{Roothaan1951}.  
In quantum chemistry, this means searching for molecular orbitals as a linear combination of atomic orbitals (MO-LCAO).
In the case of atomic matter waves, being eigenstates of a profile trap potential, each set state can be approximated by a relevant linear combination of the separated site orbitals.

The proper orbital functions can be straightforwardly constructed with the aid of the symmetry properties of the point groups $S_2\sim C_2$ (three particles) and $D_{2h}$ (four particles). Both are abelian and have only one-dimensional representations. With omitting the derivation details for three particles we obtain
\begin{eqnarray}
\phi_{g}(\mathbf{r})&=&q\,\left[\varphi_{A}(\mathbf{r})+\varphi_{B}(\mathbf{r})+\varphi_{C}(\mathbf{r})\right]%
\nonumber\\%
\phi_{e}(\mathbf{r})&\propto& qf\,\varphi_{A}(\mathbf{r})-q\,\left[\varphi_{B}(\mathbf{r})+\varphi_{C}(\mathbf{r})\right]%
\nonumber\\%
\phi_{e'}(\mathbf{r})&=& p\,\left[\varphi_{B}(\mathbf{r})-\varphi_{C}(\mathbf{r})\right]%
\label{4.2}%
\end{eqnarray}
where
\begin{eqnarray}
q&=&\frac{1}{\left(3 + 2 S_{AB}+2S_{BC}+2S_{AC}\right)^{1/2}}%
\nonumber\\%
p&=&\frac{1}{\left[2\left(1-S_{BC}\right)\right]^{1/2}}%
\nonumber\\%
f&=&\frac{2\left(1+S_{BC}\right)+S_{AB}+S_{AC}}{1+S_{AB}+S_{AC}}%
\label{4.3}%
\end{eqnarray}
and $S_{AB},\,S_{BC},\ldots$ denote the respective overlap integrals
\begin{eqnarray}
S_{AB}&=&\langle \varphi_{A}|\varphi_{B}\rangle=S_{AB}(a,h,\ldots)%
\nonumber\\%
S_{BC}&=&\langle \varphi_{B}|\varphi_{C}\rangle=S_{BC}(a,h,\ldots)%
\nonumber\\%
&&\ldots%
\label{4.4}%
\end{eqnarray}
All these integrals are assumed to be real, symmetric, and $S_{AB}=S_{AC}$. The integrals are functions of the inter-site separations and of the parameters of the original site orbitals. 

In (\ref{4.2}) we have ordered the orbitals so that the most symmetric representation with $+1$ characters for both the $C_2$-group elements, with equal superposition of the trap sites, approaches the ground eigenstate in a profile potential. The other two states with orbital energies $\epsilon_e<\epsilon_{e'}$ are ordered under the assumption that $h\gtrsim a$, see Fig.~\ref{fig3}. The skipped normalization factor in the second line of (\ref{4.2}) is expressed by a complex function of the overlap integrals.  

In the case of four particles the MO-LCAO's are given by
\begin{eqnarray}
\phi_{g}(\mathbf{r})&\propto&\varphi_{A}(\mathbf{r})+\varphi_{B}(\mathbf{r})+\varphi_{C}(\mathbf{r})+\varphi_{D}(\mathbf{r})%
\nonumber\\%
\phi_{e}(\mathbf{r})&\propto&\varphi_{A}(\mathbf{r})-\varphi_{B}(\mathbf{r})-\varphi_{C}(\mathbf{r})+\varphi_{D}(\mathbf{r})%
\nonumber\\%
\phi_{e'}(\mathbf{r})&\propto&-\varphi_{A}(\mathbf{r})-\varphi_{B}(\mathbf{r})+\varphi_{C}(\mathbf{r})+\varphi_{D}(\mathbf{r})%
\nonumber\\%
\phi_{e''}(\mathbf{r})&\propto&-\varphi_{A}(\mathbf{r})+\varphi_{B}(\mathbf{r})-\varphi_{C}(\mathbf{r})+\varphi_{D}(\mathbf{r})%
\label{4.5}%
\end{eqnarray}
where the normalization factors can be expressed as functions of the overlap integrals, defined by (\ref{4.4}), where $S_{AB}=S_{AB}(a,b,\ldots)$, $S_{BC}=S_{BC}(a,b,\ldots)$ etc., and $S_{AB}=S_{DC}$, $S_{AD}=S_{BC}$, $S_{AC}=S_{BD}$. We have ordered the orbital functions so that the most symmetric orbital, with $+1$ characters for all $D_{2h}$-group elements, belongs to the ground state $\phi_{g}(\mathbf{r})$, and $\phi_{e}(\mathbf{r})$ approaches the low energy excited orbital, i.e. $\epsilon_e<\epsilon_{e'}<\epsilon_{e''}$ if $b>a$, see Fig.~\ref{fig3}.

The suggested molecular orbitals (\ref{4.2}) and (\ref{4.5}) have been constructed as superposition of the ground state site wavefunctions -- $\varphi_{A}(\mathbf{r})$, $\varphi_{B}(\mathbf{r})$, \ldots -- of atoms, confined with the trap potentials. Then without loss of generality we have assumed them as real functions and the basic site orbitals as real and positive. 

In the orbital model the probability density function for a single particle, contributing to the collective ground state, obeys the Hund’s rule of maximum multiplicity, and for the systems of three and four particles  is respectively given by
\begin{equation}
\rho(\mathbf{r})=\frac{2}{3}\phi^2_{g}(\mathbf{r})+\frac{1}{3}\phi^2_{e}(\mathbf{r})%
\label{4.6}%
\end{equation}
and 
\begin{equation}
\rho(\mathbf{r})=\frac{1}{2}\phi^2_{g}(\mathbf{r})+\frac{1}{2}\phi^2_{e}(\mathbf{r})%
\label{4.7}%
\end{equation}
Both these distributions are independent on the total spin and give us only preliminary indicator of either charge or matter wave distribution in a particular spatial configuration.

The correlation effects, including interparticle interaction, can be followed from the density function of the joint probability distribution for two particles $\rho^{(\sigma)}(\mathbf{r}_1,\mathbf{r}_2)$, and the product $\rho(\mathbf{r}_1)\,\rho(\mathbf{r}_2)$ gives us the benchmark of independent events. The probability density $\rho^{(\sigma)}(\mathbf{r}_1,\mathbf{r}_2)$ can be extracted from the general joint density distributions evaluated in Appendix \ref{Appendix_B}. Here, the superscribed index "$\sigma$" denotes a set of spin quantum numbers, specifying the concrete coupling scheme of the spin angular momenta in (\ref{b.3})-(\ref{b.5}) (three particles) and (\ref{b.8})-(\ref{b.10}) (four particles).
 
Surprisingly, for the collective ground state, the joint variability occurs independently of the choice of spin coupling.
For the density functions of probability distributions for a pair of position variables in the system of three and four particles we respectively obtain 
\begin{eqnarray}
\lefteqn{\rho^{(0;1/2)}(\mathbf{r}_1,\mathbf{r}_2)=\rho^{(1;1/2)}(\mathbf{r}_1,\mathbf{r}_2)}%
\nonumber\\%
&=&\frac{1}{3}\left[\phi_g^2(\mathbf{r}_1)\phi_g^2(\mathbf{r}_2)+\phi_g^2(\mathbf{r}_1)\phi_e^2(\mathbf{r}_2)+\phi_e^2(\mathbf{r}_1)\phi_g^2(\mathbf{r}_2)\right.
\nonumber\\
&&\left.\phantom{\phi_g^2(\mathbf{r}_1)} - \phi_g(\mathbf{r}_1)\phi_e(\mathbf{r}_1)\phi_g(\mathbf{r}_2)\phi_e(\mathbf{r}_2)\right]%
\label{4.8}%
\end{eqnarray}
and 
\begin{eqnarray}
\lefteqn{\rho^{(0,0;0)}(\mathbf{r}_1,\mathbf{r}_2)=\rho^{(1,1;0)}(\mathbf{r}_1,\mathbf{r}_2)}%
\nonumber\\%
&=&\frac{1}{3}\left[\frac{1}{2}\phi_g^2(\mathbf{r}_1)\phi_g^2(\mathbf{r}_2)+\frac{1}{2}\phi_e^2(\mathbf{r}_1)\phi_e^2(\mathbf{r}_2)\right.
\nonumber\\%
&&\phantom{\frac{1}{2}\phi_g^2}+\phi_g^2(\mathbf{r}_1)\phi_e^2(\mathbf{r}_2)+\phi_e^2(\mathbf{r}_1)\phi_g^2(\mathbf{r}_2)%
\nonumber\\
&&\left.\phantom{\frac{1}{2}\phi_g^2(\mathbf{r}_1)} - \phi_g(\mathbf{r}_1)\phi_e(\mathbf{r}_1)\phi_g(\mathbf{r}_2)\phi_e(\mathbf{r}_2)\right]%
\label{4.9}%
\end{eqnarray}
with vanishing interference terms (\ref{b.5}) and (\ref{b.10}), such that $\rho^{(\mathrm{int})}(\mathbf{r}_1,\mathbf{r}_2)=0$ for both the cases. 

This result, as expected, contradicts the naive assumption that particle positions are independent random variables.
In particular, in the coincidence point $\mathbf{r}_1=\mathbf{r}_2\equiv\mathbf{r}$ we obtain
\begin{eqnarray}
\lefteqn{\rho^{(0;1/2)}(\mathbf{r},\mathbf{r})=\rho^{(1;1/2)}(\mathbf{r},\mathbf{r})}%
\nonumber\\%
&&=\frac{1}{3}\left[\phi_g^4(\mathbf{r}) +\phi_g^2(\mathbf{r})\phi_e^2(\mathbf{r})\right]<\rho^{2}(\mathbf{r})%
\label{4.10}%
\end{eqnarray}
and
\begin{eqnarray}
\lefteqn{\rho^{(0,0;0)}(\mathbf{r},\mathbf{r})=\rho^{(1,1;0)}(\mathbf{r},\mathbf{r})}%
\nonumber\\%
&=&\frac{1}{3}\left[\frac{1}{2}\phi_g^4(\mathbf{r})+\frac{1}{2}\phi_e^4(\mathbf{r}) +\phi_g^2(\mathbf{r})\phi_e^2(\mathbf{r})\right]<\rho^{2}(\mathbf{r})%
\nonumber\\%
\label{4.11}%
\end{eqnarray}
which is a clear signature of antibunching effect manifested in the considered mulitparticle system. Such a suppression of coincidence probability, associated with the second order interference of atomic matter waves, reduces the cross-interaction for any pair of particles and is physically provided by quantum entanglement of their spatial variables. In the case of electrons involved in a covalent bonding, the antibunching of their negative charges reduces their repulsive energy and makes the compound molecular system more stable.

\begin{figure}[tp]
\includegraphics[width=8 cm]{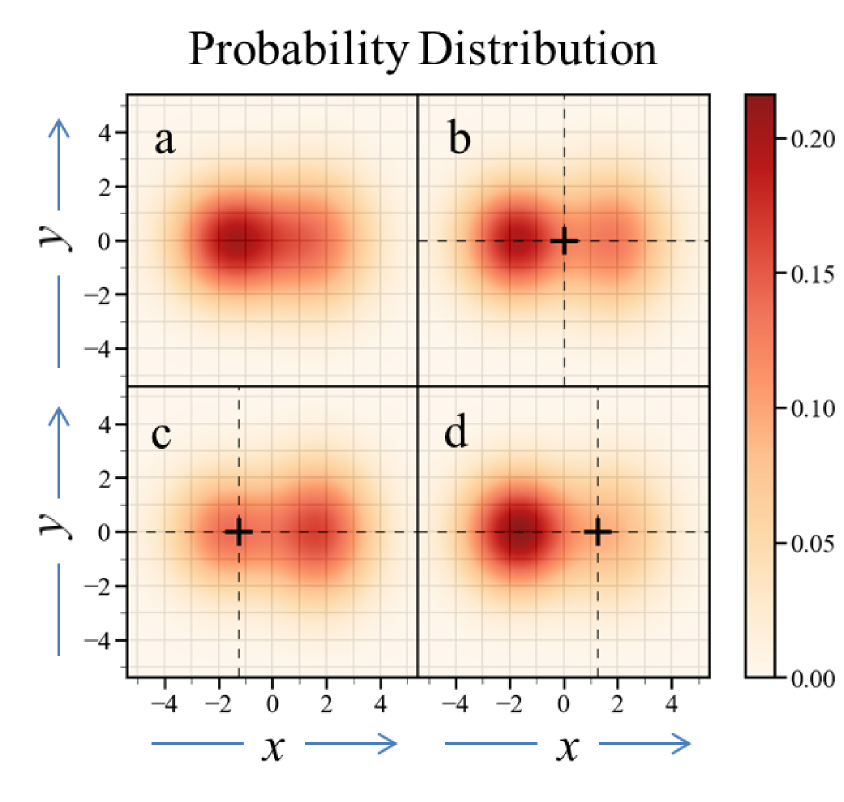}
\caption{ (a) The probability density function for a single particle, calculated for a system of three particles and triangle site configuration shown in Fig.~\ref{fig3}. Here $a=2\delta x$ and $h=2.5\delta x$, where $\delta x$ is position uncertainty of the original site state. (b)-(d) The conditional probability densities for a particle position if one particle is located at the points indicated by cross markers in the plots. The probability distributions are plotted in the $(x,y)$-plane transverse to direction of the trapping light beams with the grid unit equated to $\delta x$.}
\label{fig4}%
\end{figure}

\begin{figure}[tp]
\includegraphics[width=8 cm]{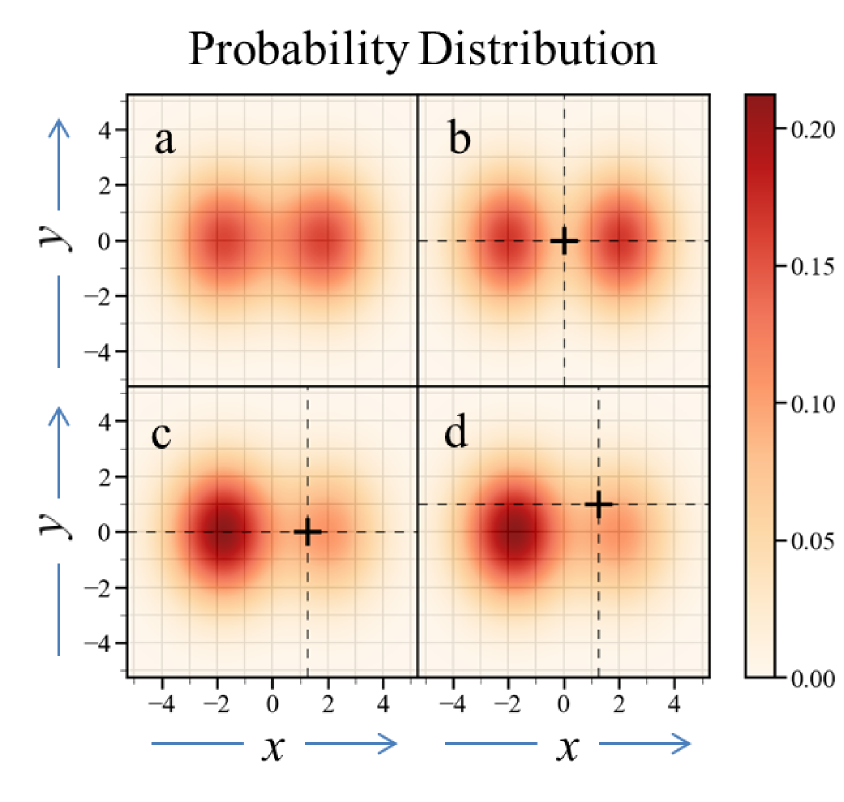}
\caption{Same as in Fig.~\ref{fig4} but for configuration of four particles, shown in Fig.~\ref{fig3}, with $a=2\delta x$ and $b=2.5\delta x$.}
\label{fig5}%
\end{figure}

This effect is visualized in the graphs of Figs.~\ref{fig4} and \ref{fig5}, where we present the probability density functions (\ref{4.6})-(\ref{4.9}) calculated for atoms confined with the microtraps for both configurations shown in Figs.~\ref{fig3}. The orbital functions (\ref{4.2}) and (\ref{4.5}) are compiled by the site wavefunctions, approximated by the ground states of the harmonic oscillators. We use Cartesian specification of the atoms' locations in the transverse $(x,y)$-plane with the $z$-axis directed along the trapping beams. The plane grid is scaled by the position uncertainties in the oscillator well $\delta x=\delta y=\sqrt{\hbar/m_a\omega_{\bot}}$, where $m_a$ and $\omega_{\bot}$ are respectively the atomic mass and frequency of transverse oscillations. In the graphs, we have plotted (a) the probability density function for a single particle $\rho(\mathbf{r})\equiv\rho(x,y)$, given by (\ref{4.6}) and (\ref{4.7}), and (b-d) the conditional probability densities $\rho^{(\sigma)}(\mathbf{r}|\mathbf{r}_0)\equiv\rho^{(\sigma)}(x,y|x_0,y_0)$, extracted from (\ref{4.8}) and (\ref{4.9}). In the latter case, the selected conditional locations $(x_0,y_0)$ of a particle are pointed by cross markers on the plots. 

The dramatic difference between the probability density functions and conditional probabilities indicates strong covariativity in the atoms' locations. The entanglement of the particle positions is a direct consequence of their quantum conjugation with their pseudospin variables. In experiment the effect of quantum antibunching can be verified by separation of the particles after interaction cycle, which should primary lead to individual occupation by one particle per each single site. That can be treated as a general manifestation of the Hong-Ou-Mandel-type interference in the system of three and four indistinguishable particles having minimal total spin.   

Let us follow the system behavior and transformation of orbital functions near the symmetric site configuration. As an example, if the rectangular configuration of four sites in Fig.~\ref{fig2} improves up to a square one, the orbital states $\phi_{e}(\mathbf{r})$ and $\phi_{e'}(\mathbf{r})$ become degenerate. We can superpose them in two alternative basis sets as
\begin{eqnarray}
&\frac{1}{\sqrt{2}}\left[\phi_{e}(\mathbf{r})\pm\phi_{e'}(\mathbf{r})\right]&%
\nonumber\\%
&\frac{1}{\sqrt{2}}\left[\phi_{e}(\mathbf{r})\pm i\phi_{e'}(\mathbf{r})\right]&%
\label{4.12}
\end{eqnarray}
These states can be occupied by repopulation of atoms with a $\pi/2$-Raman pulse on a stage of adiabatic transfer from a rectangular to square configuration. The critical feature is that, being degenerate, the orbital functions of higher symmetry can be expressed by complex-valued functions. 
The complexity creates a probability density flux when the collective stationary state possesses a persistent spatial motion.

\begin{figure}[tp]
\includegraphics[width=8 cm]{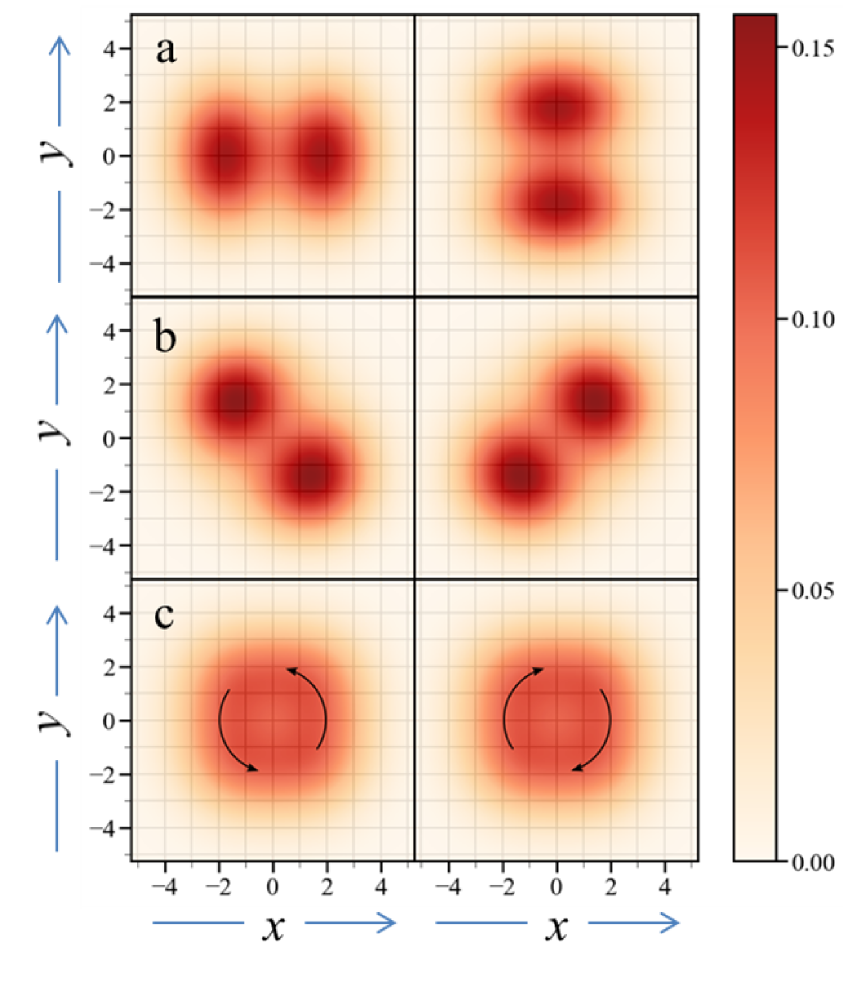}
\caption{The probability density function for a system of four particles and square-type configuration, shown in Fig.~\ref{fig3}, with $a=b=2\delta x$: (a) for occupied either $e$ (left) or $e'$ (right) orbitals; (b) for two orthogonal real-weighted superpositions of $e$ and $e'$, given by the first line in (\ref{4.12}); (c) for two orthogonal complex-valued superpositions of $e$ and $e'$, given by the second line in (\ref{4.12}). In the last case there are two opposite rotations for the probability density flux.}
\label{fig6}%
\end{figure}

To illustrate this, in Fig.~\ref{fig6} we show the density probability distributions for a square-type symmetrical configuration with subsequent occupation of different excited orbitals. In the case of complex-valued orbital, given by the second line in (\ref{4.12}), the probability density flux, visualized by plot Fig.~\ref{fig6}(c),  reveals two oppositely directed persistent rotations of the matter waves in a ring-type cavity, created by the trap centers. In this association, the occupation of real-valued orbitals, visualized in plots (a) and (b), can be treated as different realizations of the standing waves, created in the cavity by superposition of the two counterpropagating running waves.    

\subsection{Experimental insight}

\noindent The simultaneous operation by more than two atomic qubits, with overlapped matter waves, leads to more options for mediation by detection channels of the separated qubits. The interparticle interaction, initiated by overlapping the traps, gives us a resource for either internal transformation of entanglement or for its export from one to another pair of atomic qubits. From this point of view, the related experiments have a potential for preparation of various entangled states of quantum registers beyond capabilities of a quite fragile protocol of Rydberg blockade.  

As we have obtained, unlike the basic examples of Hong-Ou-Mandel interference, discussed in Section \ref{Section_II}, the antibunching in the system of three and four indistinguishable particles with minimal total spin, is insensitive as to the type of particle statistics as to the type of spin/pseudospin coupling. Nevertheless the basis states $\Psi_1^{(f/b)}(\ldots)$ and $\Psi_2^{(f/b)}(\ldots)$, parameterized by different coupling schemes, are not identical and could be resolved. These functions are orthogonal and higher order spatial and spin sensitive correlations are different for them. That would be difficult to verify for electrons, but could be done for atoms. 

The coincidence of joint distributions in (\ref{4.8}) and (\ref{4.9}) just means that both these states provide identical internal interaction with double degeneracy of the collective ground state. They lead to self-organization of a stable electron configuration. It is a natural process in chemistry and the obtained similarity with atoms allows to replicate it beyond empiric chemical models. However its simulation in experiment by the system of trapped atoms is still challenging task and would need in special technical efforts, such as protocols of Raman sideband cooling and repopulation to the system ground state by optical pumping. 

In experiment it would be interesting to follow dynamical transformation of the spatially transferred atomic qubits. Then the originally prepared system state, like (\ref{4.1}), would superpose a large number of combinatorial realizations of different position wavefunctions having same permutation symmetry. Each contributed partial state would follow to its specific temporal evolution, which  at certain critical spatial configurations would be partly disturbed by non-adiabatic transitions between the competing system states. That refers us to a crucial step of chemical transformations (either reactions or isomerizations), which, as we believe, could be reproducible by manipulation with atomic matter waves. 

\subsection{The quantum bit encoded in the degenerate ground state}

\noindent By concluding this section let us make one remark concerning an intriguing consequence of the above analysis. Namely, matter waves, expanding in space, can manifest themselves via collective effects of superfluidity or superconductivity. Here we have met both of them.

The obtained degeneracy of the ground state is an expectable consequence for a system of more than two indistinguishable and interacting particles \cite{LandauLifshitzIII}. For the basic, and slightly asymmetric, configurations, shown in Fig.~\ref{fig3}, the double degeneracy is connected with violation of "no nodes" theorem for the ground state when the number of spin one-half particles is more than two. Then both the eigenstates $\Psi_1^{(f/b)}(\ldots)$ and $\Psi_2^{(f/b)}(\ldots)$ are structured as two different combinations of the same orbital functions. Since the orbitals can be chosen as real functions, such eigenstates, having zero probability density flux, cannot be organized into quantum currents.  

However the quantum current, created by the matter waves, can optionally appear for higher symmetry states. In example of four particles and for square-type orbital symmetry, the various density distributions are shown in Fig.~\ref{fig6}. The complex-valued orbital functions lead to either right- or left-handed ring-type rotating currents, as shown in Fig.~\ref{fig6}$(c)$. Alternatively we can imagine two standing-type orthogonal real-valued realizations of $\Psi_1^{(f/b)}(\ldots)$ and $\Psi_2^{(f/b)}(\ldots)$, as superpositions of two counterrotating waves, as shown in Fig.~\ref{fig5}$(a,b)$. For a fermionic system we foresee here certain similarity with Josephson qubits created by superconductive currents in macroscopic metallic samples. However it makes difference that the considered system is not described by a single function of order parameter, but instead is characterized by superposition of various position wavefunctions, see Appendix \ref{Appendix_A}. 

In organic chemistry existence of laser-induced electron currents in chiral aromatic molecules were earlier predicted in \cite{Kanno2010}. For an electronic cluster, structured into a divalent bond, it means a natural realization of the superconductive quantum bit, which could be utilized as a physical resource for the information mapping and processing in organic molecule. The system seems as scalable and the obtained degeneracy can be straightforwardly generalized up to multivalent bonds, normally attributing any such molecule. This subtle issue, concerning the background of biological microprocessor, is far beyond the scope of our paper. But we can point out that with atomic matter waves we could imitate a possible data processing by preparation of a controllable superposition of the above basis states. It seems quite feasible for the modern state-of-the-art techniques to organize a quantum bus among several such mesoscopicaly scaled superfluid quantum bits by their coupling via external optical channels. 

\section{Conclusion}

\noindent In the paper, we have shown that in some specific physical conditions the indistinguishable particles, existing in the same environment, can similarly behave independently on their quantum statistics. 
The main requirement for that is to operate with the systems consisting up to four particles having minimal collective spin angular momentum or collective pseudospin. Then the relevant quantum conjugation of the pseudospin and position variables tends the bosonic atoms to behave as fermions.

Under these conditions, an array of bosonic atoms, mediated by optical tweezers and localized on a mesoscopic scale, could be implemented for studying a fermionic system of interacting electrons. The electrons are assumed to be localized in a nanoscale area, with the electrons providing covalent bonding in a molecule. For enclaves consisting of up to four electrons, the molecular monovalent and divalent bonding could be realistically reproduced by matter waves of neutral atoms confined with optical microtraps emulating the nuclear centers. 

In our classification of the collective quantum states, we have focused on the examples where the effects of second-order interference and position entanglement in many-particle systems would be meaningful.  In particular, the obtained antibunching of the particle positions, existing for electrons, can be replicated by atoms. Antibunching reduces electron repulsion and can play a supporting role in enforcing divalent bonds in molecules. The antibunching can be verified in an experiment with atomic matter waves by the controllable separation of trapping sites in a generalized Hong-Ou-Mandel detection protocol. 

In perspective, the analogue quantum simulations of dynamical transformations in multiparticle systems could be implemented to follow chemical reactions or isomerization processes. The dynamical behavior of compound objects would be quite difficult to follow by classical computational algorithms if the number of interacting parties is more than two. However, it seems to be a realistic task for the proposed quantum simulator, where different reaction paths can be replicated with spatial transfer of atoms by optical tweezers. For example, it could be feasible to simulate a non-adiabatic transition between close energy surfaces approaching each other at certain critical spatial configurations. 

The obtained degeneracy of the ground state provides a physical resource for either superconductive (electrons) or superfluid (atoms) dynamics. The key difference with the well-known macroscopic description of such quantum phenomena is that in the considered physical realization the quantum current is characterized by several position wavefunctions not reducible to a single order parameter. This is a direct consequence of the two-dimensionality of the symmetric group representation for the considered spin and pseudospin configurations. In the case of electrons and big molecules we foresee here a certain physical option for self-organization of a memory unit, incorporated in an entire information processing in organic systems. We believe that such a physical scenario of information processing in molecules could be simulated by mesoscopically scaled superfluid atomic quantum bits by their coupling via external optical channels. 

\section*{Acknowledgments}
\noindent This work was supported by the Russian Science Foundation under Grant No. 23-72-10012, and by Rosatom in framework of the Roadmap for Quantum computing (Contracts No. 868-1.3-15/15-2021 and No. P2154). 
A.M. acknowledges the project CZ.02.01.01/00/22\_008/0004649 (QUEENTEC) of EU and MEYS Czech Republic and the project GA23-06224S of the Czech
Science Foundation.
L.V.G. and N.A.M. acknowledge support from the Foundation for the Advancement of Theoretical Physics and Mathematics “BASIS” under Grant No. 23-1-2-37-1.

\appendix

\section{Multiparticle fermion system}\label{Appendix_A}

\noindent In this section we briefly overview our basic chemical object, namely, the system of many electrons/fermions confined with the profile potential and interacting each other. We assume that the internal Hamiltonian is spin independent, in its non-relativistic approximation, and that makes highly effective studying of the eigenstate problem by the methods of the group theory. We make use of that the symmetric group strongly mediates the properties of the energy eigenstates.  By aiming to simplify the math formalism we restrict our discussion by the physical systems consisting of two, three and four electrons. For a sake of notation clarity we  specify the spin state vectors in Dirac notation, and the position state vectors by functions $\Phi(i,j,k,\ldots)$, where $i,j,k,\ldots$ are the integer numbers enumerating the position coordinates $\mathbf{r}_i,\mathbf{r}_j,\mathbf{r}_k,\ldots$. In the end of the section we turn to the derivation consequences linking with the similar systems of bosonic atoms having one-half pseudospins.

\subsection{Two electrons}

\noindent For two interacting electrons the entire wave-function of each energy state is factorized in the product of the collective spin and the position wave-functions 
\begin{eqnarray}
\Psi^{(0)}(1,2)&=&|0,0\rangle\,\Phi^{(0)}(1,2)%
\nonumber\\%
\Psi^{(1)}_{M}(1,2)&=&|1,M\rangle\,\Phi^{(1)}(1,2)%
\label{a.1}%
\end{eqnarray}
where $M$ is the projection of the total spin $S=0,1$, which is $M=0$ for the singlet state. For interacting particles the functions $\Phi^{(0)}(1,2)$ and $\Phi^{(1)}(1,2)$ belong to different energies should obey the permutation rules
\begin{eqnarray}
\Phi^{(0)}(1,2)&=&\Phi^{(0)}(2,1)%
\nonumber\\%
\Phi^{(1)}(1,2)&=&-\Phi^{(1)}(2,1)%
\label{a.2}%
\end{eqnarray}
This defines the simplest one-dimensional representations for the symmetric group for two objects with only one permutation. It clear shows that the energy states of the Hamiltonian are specified by this permutation symmetry. In the main text we have verified the triplet/singlet conservation symmetry via the Hong-Ou-Mandel effect for independent fermions, but here we extend this effect beyond the self-consistent orbital approximation.

Intuitively, the position bunching of electrons in the singlet state lies in background of the monovalent bonding in chemistry. In general for an electron pair in a singlet state the density of negative charge distribution is given by 
\begin{equation}
n(\mathbf{r})=2\,\rho^{(0)}(\mathbf{r})=2\int d^3r' |\Phi^{(0)}(\mathbf{r},\mathbf{r}')|^2%
\label{a.3}%
\end{equation}
and can be thought as enhanced (by integration near $\mathbf{r}'\sim\mathbf{r}$) in the area between the positively charged nuclei and provide more stable configuration for the compound system. But, as clarified in the main text, this point should be revised in a maltiparticle system.

\subsection{Three electrons}
\noindent For the system consisting of three interacting electrons the permutation group is non-abelian and we can refer to general statements of the group theory for the symmetric group $S_N$ of $N$ elements.  There are the following requirements to any stationary solution of the Schr\"{o}dinger equation: (i) The solution has to be factorized to the product of the spin and position wavefunctions; (1) The Hamiltonian commutes with any of $N!$ permutations from the $S_{N}$-group -- $[\hat{H},P]=0$ -- where $N$ is the number of electrons, and, as a consequence, the spin and position component, considered independently, should belong to a certain irreducible representation of this group; (iii) In accordance with the fermionic statistics the complete wavefunction should be given only by antisymmetric one-dimensional representation.

In our example the $S_3$-group contains six elements. Then the dimensions of representations are restricted by the group order and obey the condition: $1^2+1^2+2^2=6$. So there are two one-dimensional representations, which we denote as $\hat{\mathbf{D}}^{(1)}\equiv\hat{\mathbf{S}}$ (symmetric) and $\hat{\tilde{\mathbf{D}}}^{(1)}\equiv\hat{\mathbf{A}}$ (antisymmetric) and one two-dimensional $\hat{\mathbf{D}}^{(2)}\sim\hat{\tilde{\mathbf{D}}}^{(2)}$. Here and throughout by tilde we denote the representations, associated with the transpose partition of $N$, and expressed by the Young diagram reflected along the main diagonal. Then the two-dimensional representation can be expressed in two equivalent forms. In the spin subspace it can be realized only one dimensional $\hat{\mathbf{D}}^{(1)}$ (for the maximal spin $S=3/2$) and two-dimensional (for various realizations of $S=1/2$). In the position subspace all three representations are accessible for construction of the different energy states.

As proven in the group theory applied to quantum mechanics, to construct the complete wavefunction the direct product of two irreducible representations of the same rank $\hat{\mathbf{D}}^{(K)}\otimes\hat{\tilde{\mathbf{D}}}^{(K)}$, respectively acting in the spin and position subspaces, has to be decomposed in the direct sum of irreducible representations, see \cite{Messiah1961}. There is only one antisymmetric representation $\hat{\tilde{\mathbf{D}}}^{(1)}\equiv \hat{\mathbf{A}}$ contributed to this product. However the alternative choice in the product order $\hat{\tilde{\mathbf{D}}}^{(K)}\otimes\hat{\mathbf{D}}^{(K)}$ generates different antisymmetric state. In a collection of spin one-half particles (not only fermions but atomic bosons as well) the representation $\hat{\mathbf{D}}^{(K)}$ of rank $K$ is uniquely determined by the total spin $S$ and constructively built up by the angular momentum algebra.

Based on requirements (i)--(iii), the representation $\hat{\mathbf{D}}^{(2)}\otimes\hat{\tilde{\mathbf{D}}}^{(2)}$ suggests the following energy eigenstate for the total spin $S=1/2$:
\begin{eqnarray}
\Psi^{(0;1/2)}_M(1,2,3)&=&|s_1,s_{23}=0;1/2,M\rangle\;\Phi^{(0;1/2)}(1;2,3)%
\nonumber\\%
&+&|s_2,s_{31}=0;1/2,M\rangle\;\Phi^{(0;1/2)}(2;3,1)%
\nonumber\\%
&+&|s_3,s_{12}=0;1/2,M\rangle\;\Phi^{(0;1/2)}(3;1,2)%
\nonumber\\%
\label{a.4}%
\end{eqnarray}
Here and in other expression in this section, where it is important, we clarify in the ket- and bra-vectors the concrete values for the intermediate spin coupling. So in (\ref{a.4}) $s_{jk}=0$ stands for the intermediate coupling of $jk$-pair to a singlet state and the spin of the third particle $s_i=1/2$, with $i\neq j\neq k$, equates to the total spin. We superscribe and specify the complete wavefunction $\Psi^{(0;1/2)}_M(1,2,3)$ as well as the position wavefunction $\Phi^{(0;1/2)}(i;j,k)$ by the quantum numbers of intermediate coupling $s_{jk}=0$, total spin $S=1/2$ and its projection $M=\pm 1/2$. The position  wavefunction $\Phi^{(0;1/2)}(i;j,k)$ is symmetric in respect to permutation of the second pair of arguments $j\rightleftarrows k$ such that
\begin{equation}
\Phi^{(0;1/2)}(i;j,k)=\Phi^{(0;1/2)}(i;k,j)%
\label{a.5}
\end{equation}
There are three functions, fulfilling (\ref{a.5}), but they should not be treated as linearly independent and their meaningful choice is insensitive to the sum
\begin{equation}
\Phi^{(0;1/2)}(1;2,3)+\Phi^{(0;1/2)}(2;3,1)+\Phi^{(0;1/2)}(3;1,2)%
\label{a.6}
\end{equation}
that is ultimately excluded by expansion (\ref{a.4}). So only two of these three functions constitute the group representation in the position subspace, and we can additionally set 
\begin{equation}
\Phi^{(0;1/2)}(1;2,3)+\Phi^{(0;1/2)}(2;3,1)+\Phi^{(0;1/2)}(3;1,2)=0%
\label{a.7}
\end{equation}
accordingly to two-dimensionality of the representation.

By alternative order in the representation product $\hat{\tilde{\mathbf{D}}}^{(2)}\otimes\hat{\mathbf{D}}^{(2)}$ we obtain
\begin{eqnarray}
\Psi^{(1;1/2)}_M(1,2,3)&=&|s_1,s_{23}=1;1/2,M\rangle\;\Phi^{(1;1/2)}(1;2,3)%
\nonumber\\%
&+&|s_2,s_{31}=1;1/2,M\rangle\;\Phi^{(1;1/2)}(2;3,1)%
\nonumber\\%
&+&|s_3,s_{12}=1;1/2,M\rangle\;\Phi^{(1;1/2)}(3;1,2)%
\nonumber\\%
\label{a.8}%
\end{eqnarray}
where the each position wavefunction is now antisymmetric in respect to permutation of the second pair of arguments $j\rightleftarrows k$ and
\begin{equation}
\Phi^{(1;1/2)}(i;j,k)=-\Phi^{(1;1/2)}(i;k,j)%
\label{a.9}
\end{equation}
Here unlike (\ref{a.4}) the electron pair $j,k$ is coupled to the intermediate triplet spin state $s_{jk}=1$. Again these functions are not linearly independent since the composition (\ref{a.8}) is insensitive to the sum
\begin{equation}
\Phi^{(1;1/2)}(1;2,3)+\Phi^{(1;1/2)}(2;3,1)+\Phi^{(1;1/2)}(3;1,2)%
\label{a.10}
\end{equation}
and we can set
\begin{equation}
\Phi^{(1;1/2)}(1;2,3)+\Phi^{(1;1/2)}(2;3,1)+\Phi^{(1;1/2)}(3;1,2)=0%
\label{a.11}
\end{equation}
So, again, we obtain that only two of these functions represent the symmetric group in the position subspace. Note that there are two different sets of two linearly independent position wavefunctions given by mathematical solution of the non-relativistic Schr\"{o}dinger equation, which, with cyclic permutation of their arguments, contribute in (\ref{a.4}) and (\ref{a.8}) .

The basis spin states, contributing in (\ref{a.4}) and (\ref{a.8}) are not mutually orthogonal and not linearly independent. Being components of two-dimensional representations they fulfill the identity 
\begin{equation}
|s_1,s_{23};1/2,M\rangle+|s_2,s_{31};1/2,M\rangle+|s_3,s_{12};1/2,M\rangle=0%
\label{a.12}%
\end{equation}
valid for $s_{23}=s_{31}=s_{12}$. This can be verified by subsequent projecting the left-hand side on each of the contributing spin state. The overlapping matrix elements are expressed Wigner's 6j-simbols, see \cite{Varshalovich1988}. As example,
\begin{eqnarray}
\lefteqn{1+\langle s_1,s_{23};1/2,M|s_2,s_{31}=s_{23};1/2,M\rangle}%
\nonumber\\%
&&+\langle s_1,s_{23};1/2,M|s_3,s_{12}=s_{23};1/2,M\rangle
\nonumber\\%
\nonumber\\%
&=&1+(-)^{s_{23}}2(2s_{23}+1)\left\{\begin{array}{ccc}1/2&1/2&s_{23}\\ 1/2&1/2&s_{23}\end{array}\right\}=0%
\nonumber\\%
\label{a.13}%
\end{eqnarray}
which is valid for both $s_{23}=0,1$.

The constructed states (\ref{a.4}) and (\ref{a.8}) are mutually orthogonal. Their scalar product is given by
\begin{widetext}
\begin{eqnarray}
\lefteqn{\langle\Psi^{(0;1/2)}_M(1,2,3)|\Psi^{(1;1/2)}_M(1,2,3)\rangle} 
\nonumber\\%
&=& \langle s_1,s_{23}=0;1/2,M|s_2,s_{31}=1;1/2,M\rangle\int d(1)d(2)d(3)\;\Phi^{(0;1/2)\ast}(1;2,3)\cdot\Phi^{(1;1/2)}(2;3,1)%
\nonumber\\%
&& +\langle s_1,s_{23}=0;1/2,M|s_3,s_{12}=1;1/2,M\rangle\int d(1)d(2)d(3)\;\Phi^{(0;1/2)\ast}(1;2,3)\cdot\Phi^{(1;1/2)}(3;1,2)+\ldots%
\nonumber\\%
\nonumber\\%
&=&\sqrt{3}\left\{\begin{array}{ccc}1/2&1/2&1\\ 1/2&1/2&0\end{array}\right\}\int d(1)d(2)d(3)\;\Phi^{(0;1/2)\ast}(1;2,3)\left[\Phi^{(1;1/2)}(2;3,1) + \Phi^{(1;1/2)}(3;1,2)\right] +\ldots%
\label{a.14}%
\end{eqnarray}
\end{widetext}
where the ellipses denote other terms generated by cyclic permutations of $1,2,3$ in the wavefunction arguments in a such way that in the products of functions the argument orders are always different for the multipliers. We make use of identity (\ref{a.11}) and then obtain
\begin{eqnarray}
\lefteqn{\langle\Psi^{(0;1/2)}_M(1,2,3)|\Psi^{(1;1/2)}_M(1,2,3)\rangle}%
\nonumber\\%
&=&-\frac{\sqrt{3}}{2}\int d(1)d(2)d(3)\;\Phi^{(0;1/2)\ast}(1;2,3)\cdot\Phi^{(1;1/2)}(1;2,3)%
\nonumber\\%
&&+\ldots=0%
\label{a.15}
\end{eqnarray}
where the position integrals vanish because of either symmetry (\ref{a.5}) or antisymmetry (\ref{a.9}) to the argument transposition.

The third antisymmetric representation is created by the direct product $\hat{\mathbf{D}}^{(1)}\otimes\hat{\tilde{\mathbf{D}}}^{(1)}$ and reveals the following ferromagnetic state 
\begin{equation}
\Psi^{(1;3/2)}_{M}(1,2,3)=|3/2,M\rangle\,\Phi^{(1;3/2)}(1,2,3)%
\label{a.16}%
\end{equation}
where the spin state $|S,M\rangle$, with the total spin $S=3/2$, is symmetric and the position wavefunction $\Phi^{(1;3/2)}(1,2,3)$ is antisymmetric in respect to any permutation of its arguments.

\subsection{Four electrons}

\noindent The $S_4$-group contains twenty four elements. The dimensions of representations are constrained by the group order and obey the condition: $1^2+1^2+2^2+3^2+3^2=24$. But for the spin one-half fermions there are applicable only one one-dimensional, one two-dimensional (in two equivalent forms), and one three-dimensional representations. In these cases the total spin $S$ takes three values and is respectively connected with the preparation of $S=3$ (one ferromagnetic state with maximized spin), $S=0$ (two singlet anti-ferromagnetic states) and $S=1$ (one triplet state). Let us be focused here only on the singlet states, which, as we further show, are constructed similarly to the states (\ref{a.4}) and (\ref{a.8}). 

First singlet state $S=M=0$ contributes to the representation product $\hat{\mathbf{D}}^{(2)}\otimes\hat{\tilde{\mathbf{D}}}^{(2)}$ and is expressed by superposition of three different spin pairs, already coupled to the singlet states. It is given by
\begin{eqnarray}
\Psi^{(0,0;0)}(1,2,3,4)&=&|s_{12},s_{34}=0;00\rangle\;\Phi^{(0,0;0)}(1,2;3,4)%
\nonumber\\%
&+&|s_{23},s_{14}=0;00\rangle\;\Phi^{(0,0;0)}(2,3;1,4)%
\nonumber\\%
&+&|s_{31},s_{24}=0;00\rangle\;\Phi^{(0,0;0)}(3,1;2,4)%
\nonumber\\%
\label{a.17}%
\end{eqnarray}
where we have slightly modified and simplified our notations, and omitted the specification of individual spins $s_1=s_2=s_3=s_4=1/2$. For the intermediate coupled spins we clarify their values: as example, here $s_{ij},s_{kl}=0$ means $s_{ij}=0$ and $s_{kl}=0$ and similarly in other expressions in this section. Since the three contributed spin states belong to  $\hat{\mathbf{D}}^{(2)}$ representation only two of them are linearly independent, such that
\begin{equation}
|s_{12},s_{34}=0;00\rangle+|s_{23},s_{14}=0;00\rangle+|s_{31},s_{24}=0;00\rangle=0%
\label{a.18}%
\end{equation}
In turn the position wavefunction $\Phi^{(0,0;0)}(i,j;k,l)$ is symmetric in respect to independent transpositions $i\rightleftarrows j$ and $k\rightleftarrows l$. Additionally this function is symmetric to the pair permutations $i,j\rightleftarrows k,l$ such that
\begin{eqnarray}
\Phi^{(0,0;0)}(i,j;k,l)&=&\Phi^{(0,0;0)}(j,i;k,l)%
\nonumber\\%
\Phi^{(0,0;0)}(i,j;k,l)&=&\Phi^{(0,0;0)}(i,j;l,k)%
\nonumber\\%
\Phi^{(0,0;0)}(i,j;k,l)&=&\Phi^{(0,0;0)}(k,l;i,j)%
\label{a.19}
\end{eqnarray}
The superposition (\ref{a.17}) eliminates the sum of the position wavefunctions and let only two them be linearly independent. So we can additionally set
\begin{equation}
\Phi^{(0,0;0)}(1,2;3,4)+\Phi^{(0,0;0)}(2,3;1,4)+\Phi^{(0,0;0)}(3,1;2,4)=0%
\label{a.20}
\end{equation}
These symmetry properties are similar to (\ref{a.9})-(\ref{a.11}) in the case of three particle.

The second singlet state $S=M=0$ contributes to the product $\hat{\tilde{\mathbf{D}}}^{(2)}\otimes\hat{\mathbf{D}}^{(2)}$  and is given by the following superposition
\begin{eqnarray}
\Psi^{(1,1;0)}(1,2,3,4)&=&|s_{12},s_{34}=1;00\rangle\;\Phi^{(1,1;0)}(1,2;3,4)%
\nonumber\\%
&+&|s_{23},s_{14}=1;00\rangle\;\Phi^{(1,1;0)}(2,3;1,4)%
\nonumber\\%
&+&|s_{31},s_{24}=1;00\rangle\;\Phi^{(1,1;0)}(3,1;2,4)%
\nonumber\\%
\label{a.21}%
\end{eqnarray}
where the particle spins are summed to the total zero angular momentum via intermediate coupling of the particle pairs to the triplet states. Similarly to (\ref{a.18}) these basis states fulfil the condition
\begin{equation}
|s_{12},s_{34}=1;00\rangle+|s_{23},s_{14}=1;00\rangle+|s_{31},s_{24}=1;00\rangle=0%
\label{a.22}%
\end{equation}
The properties (\ref{a.18}) and (\ref{a.22}) indicate their belonging to two two-dimensional representation and can be proven similarly to (\ref{a.13}), once the overlapping matrix elements are expressed by 9j-symbols, see \cite{Varshalovich1988}.

The position wavefunctions $\Phi^{(1,1;0)}(i,j;k,l)$ are antisymmetric in respect to independent transpositions $i\rightleftarrows j$ and $k\rightleftarrows l$, but symmetric to the pair permutations $i,j\rightleftarrows k,l$ such that
\begin{eqnarray}
\Phi^{(1,1;0)}(i,j;k,l)&=&-\Phi^{(1,1;0)}(j,i;k,l)%
\nonumber\\%
\Phi^{(1,1;0)}(i,j;k,l)&=&-\Phi^{(1,1;0)}(i,j;l,k)%
\nonumber\\%
\Phi^{(1,1;0)}(i,j;k,l)&=&\Phi^{(1,1;0)}(k,l;i,j)%
\label{a.23}
\end{eqnarray}
The superposition (\ref{a.21}) eliminates the sum of the position wavefunctions and only two functions are linearly independent. So we can additionally set
\begin{equation}
\Phi^{(1,1;0)}(1,2;3,4)+\Phi^{(1,1;0)}(2,3;1,4)+\Phi^{(1,1;0)}(3,1;2,4)=0%
\label{a.24}
\end{equation}
The states (\ref{a.17}) and (\ref{a.21}) are orthogonal, which can be similarly proven as for the states (\ref{a.4}) and (\ref{a.8}).

\subsection{Visualization of the interacting fermions by atomic bosons having one-half pseudospins}

\noindent The results of the above sections show us that the quantum states with minimal total spin $S=1/2$ (three particles) and $S=0$ (four particles) are similarly constructed. To stress this circumstance and go from there we  further discuss both these configurations together and denote the wavenctions for states (\ref{a.4}) and (\ref{a.17}) as $\Psi_{1}^{(f)}$, and for states (\ref{a.8}) and (\ref{a.21}) as $\Psi_{2}^{(f)}$. 

Without interparticle interaction, or if it is self-consistently incorporated to the orbital concept under Hartree-Fock method, these are degenerate states belonging to the energy of occupied orbitals. But in more general case, beyond the orbital concept, we could search the complete wavefunction as linear superposition
\begin{equation}
\Psi(\ldots)=C_1\,\Psi_{1}^{(f)}(\ldots) + C_2\,\Psi_{2}^{(f)}(\ldots)%
\label{a.25}%
\end{equation}
where ellipses denote the respective arguments. The expansion coefficients depend on interparticle interaction and on positions of the confining nuclear centers. 

The analogue quantum simulator would let us emulate the correct charge density distribution in the molecule via substitution of electrons by atoms.  From the first sight that would be impossible to do because of the exclusive difference in their quantum statistics. However the situation becomes not so undoubtable at least for the considered systems consisting of two, three and four particles. For two particles that is because of trivial correspondence between position and spin wavefunctions. As we have shown in Section \ref{Section_III}, that let us link the Hong-Ou-Mandel bunching of atomic bosons with bunching of electrons and hence with construction of a monovalent bond in molecules.

Surprisingly but the fermionic states, minimizing the total spin in collection of three and four electrons, can be relevantly reproduced by bosonic systems as well. Indeed, such states are described by two-dimensional representations of the $S_3$ and $S_4$ symmetric groups which possess by equivalent representations, expressed by the transposed Young diagrams, such that $\hat{\mathbf{D}}^{(2)}\sim\hat{\tilde{\mathbf{D}}}^{(2)}$. To construct the collective wavefunctions of bosonic atoms, having one-half pseudospins, we only need to replace $\Phi^{(0;1/2)}(i;j,k)$ by $\Phi^{(1;1/2)}(i;j,k)$ in Eq.~(\ref{a.4}) and vice versa in (\ref{a.8}): $\Phi^{(0;1/2)}(i;j,k)\rightleftarrows \Phi^{(1;1/2)}(i;j,k)$. Similarly we have to interchange $\Phi^{(0,0;0)}(i,j;k,l)\rightleftarrows \Phi^{(1,1;0)}(i,j;k,l)$ in (\ref{a.17}) and (\ref{a.21}). So instead of (\ref{a.25}) we can suggest the wavefunction
\begin{equation}
\Psi(\ldots)=C_1\,\Psi_{1}^{(b)}(\ldots) + C_2\,\Psi_{2}^{(b)}(\ldots)%
\label{a.26}%
\end{equation}
where functions $\Psi_{1}^{(b)}$ and $\Psi_{2}^{(b)}$ are expressed by conjugation of such replaced position wavefunctions with the respective singlet pseudospin wavefunctions. Although the individual pseudospin states $|a\rangle$ and $|b\rangle$, defined in Section \ref{Section_III}, have different internal energies their superposition to a total spin state has a fixed energy. 

It is crucially important that just the states having minimal spin are mainly responsible for formation of covalent bonding in chemistry. Then, as a consequence of the above derivation, for the systems of three and four particles any related joint charge distribution of electrons could be realistically reproduced by the joint matter distribution of neutral atoms. That provides an option for quantum simulations of mono and divalent bonds by manipulation up to four atoms confined with optical traps. And even more, we can follow a nonadiabatic conversion of such a prepared artificial molecular cluster between its bounded modes or decoupled channels via its scanning near certain critical configurations. The latter should exist near anti-intersection points of the energy surfaces belonging to different states of the same symmetry.

Visualization of fermionic system by bosons for the states with minimal total spin is a unique option only for three and four particles, since for symmetric group $S_N$, with $N\geq 5$, the regular and transposed representations are nonequivalent and differently conjugate the spin (pseudospin) and spatial variables.

\section{Joint probability function of the position variables}\label{Appendix_B}

\subsection{Elimination of the spin variables}

\noindent In non-relativistic approach a molecular neutral charge system is stabilized by redistribution of a electronic charge taken from valent shells of interacting atoms. As we have seen for certain  configurations the charge distribution can be displayed by neutral atoms mediated by microtraps. By varying the trap configurations we can construct a more or less realistic replica of molecular bonds by atomic matter wave density. To find the joint density distribution of particle positions we have to trace the density matrices, associated with the wavefunctions (\ref{a.25}) and (\ref{a.26}), either over spin or pseudospin variables. Here we present the results explicitly for three and four particles.

\subsubsection*{Three particles}

\noindent For fermions and bosons we get respectively
\begin{eqnarray}
\rho^{(f)}(1,2,3)&=&|C_1|^2\,\rho^{(0;1/2)}(1,2,3)%
\nonumber\\%
&&+|C_2|^2\,\rho^{(1;1/2)}(1,2,3)%
\nonumber\\%
&&+\left[C_1C_2^{\ast}\,\rho^{(\mathrm{int})}(1,2,3) + c.c.\right]%
\label{b.1}%
\end{eqnarray}
and 
\begin{eqnarray}
\rho^{(b)}(1,2,3)&=&|C_1|^2\,\rho^{(1;1/2)}(1,2,3)%
\nonumber\\%
&&+|C_2|^2\,\rho^{(0;1/2)}(1,2,3)%
\nonumber\\%
&&+\left[C_1C_2^{\ast}\,\rho^{(\mathrm{int})\ast}(1,2,3) + c.c.\right]%
\label{b.2}%
\end{eqnarray}
where 
\begin{equation}
\rho^{(0;1/2)}(1,2,3)=\frac{3}{2}\sum_{\circlearrowleft}\left|\Phi^{(0;1/2)}(i;j,k)\right|^2%
\label{b.3}%
\end{equation}
and
\begin{equation}
\rho^{(1;1/2)}(1,2,3)=\frac{3}{2}\sum_{\circlearrowleft}\left|\Phi^{(1;1/2)}(i;j,k)\right|^2%
\label{b.4}%
\end{equation}
and
\begin{eqnarray}
\lefteqn{\rho^{(\mathrm{int})}(1,2,3)}%
\nonumber\\%
&&=-\frac{\sqrt{3}}{2}\sum_{\circlearrowleft}\Phi^{(0;1/2)}(i;j,k)\cdot\Phi^{(1;1/2)\ast}(i;j,k)%
\nonumber\\%
\label{b.5}%
\end{eqnarray}
where $\circlearrowleft$ denotes the sum other three possible cyclic permutations $(1,2,3)\to (i,j,k)$. 

\subsubsection*{Four particles}

\noindent Here we similarly obtain
\begin{eqnarray}
\rho^{(f)}(1,2,3,4)&=&|C_1|^2\,\rho^{(0,0;0)}(1,2,3,4)%
\nonumber\\%
&&+|C_2|^2\,\rho^{(1,1;0)}(1,2,3,4)%
\nonumber\\%
&&+\left[C_1C_2^{\ast}\,\rho^{(\mathrm{int})}(1,2,3,4) + c.c.\right]%
\label{b.6}%
\end{eqnarray}
and 
\begin{eqnarray}
\rho^{(b)}(1,2,3,4)&=&|C_1|^2\,\rho^{(1,1;0)}(1,2,3,4)%
\nonumber\\%
&&+|C_2|^2\,\rho^{(0,0;0)}(1,2,3,4)%
\nonumber\\%
&&+\left[C_1C_2^{\ast}\,\rho^{(\mathrm{int})\ast}(1,2,3,4) + c.c.\right]%
\label{b.7}%
\end{eqnarray}
where 
\begin{equation}
\rho^{(0,0;0)}(1,2,3,4)=\frac{3}{2}\sum_{\circlearrowleft}\left|\Phi^{(0,0;0)}(i,j;k,4)\right|^2%
\label{b.8}%
\end{equation}
and
\begin{equation}
\rho^{(1,1;0)}(1,2,3,4)=\frac{3}{2}\sum_{\circlearrowleft}\left|\Phi^{(1,1;0)}(i,j;k,4)\right|^2%
\label{b.9}%
\end{equation}
and
\begin{eqnarray}
\lefteqn{\rho^{(\mathrm{int})}(1,2,3,4)}%
\nonumber\\%
&&=-\frac{\sqrt{3}}{2}\sum_{\circlearrowleft}\Phi^{(0,0;0)}(i,j;k,4)\cdot\Phi^{(1,1;0)\ast}(i,j;k,4)%
\nonumber\\%
\label{b.10}%
\end{eqnarray}
where $\circlearrowleft$ denotes the sum other three possible cyclic permutations $(1,2,3)\to (i,j,k)$. Note that because of the symmetry relations (\ref{a.19}) and (\ref{a.23}) there is no need to permutate one arbitrary selected particle. In our example that is particle \#4.

\vspace{\baselineskip}
Both the cases of three and four particles belong to identical representations (similar group classes) that explains the coincidence of the algebraic factors. The above expressions explicitly show that the difference between the position distributions for two alternative statistics is in the interchange $C_1\rightleftarrows C_2$ and it completely vanishes at the balance condition $C_1=C_2^{\ast}$. As we have commented in the main text these expansion coefficients are not arbitrary parameters of the theory and depend on the entire dynamics, geometry of reaction paths etc.. A general chemical process can be mapped and tracked via manipulation with atomic matter waves that is a strategic goal of analogue quantum simulations. It fairly reproduces all the critical features of the evolution process. As example, the joint position distribution always drops to zero at the coincidence point for more than two particles even in the case of bosons.

\subsection{The position wavefunctions approximated by molecular orbitals}

\noindent In general case it would be difficult to solve the Schr\"{o}dinger equation for many particle problem. Instead, with accepting the common quantum chemistry concept, the eigenstates can be approximated by superposed products of single particle orbitals. Each orbital is a solution of the Schr\"{o}dinger equation reduced for a single particle driven by an effective self-consistent potential. Then the orbitals, contributed to construction of any collective state, can be associated with the cells of Young diagrams. Specifically in the considered case to build up the three and four particle states we respectively define  
\begin{equation}
\phi_{\mathrm{i}}(\_)\,\phi_{\mathrm{ii}}(\_)\,\phi_{\mathrm{iii}}(\_)\Leftrightarrow\vcenter{\hbox{\includegraphics[width=2 cm]{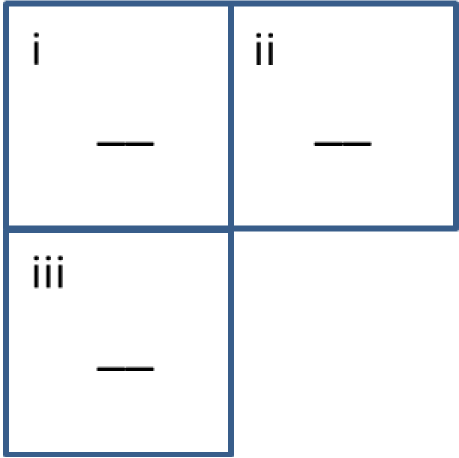}}}%
\label{b.11}%
\end{equation}
and
\begin{equation}
\phi_{\mathrm{i}}(\_)\,\phi_{\mathrm{ii}}(\_)\,\phi_{\mathrm{iii}}(\_)\,\phi_{\mathrm{iv}}(\_)\Leftrightarrow\vcenter{\hbox{\includegraphics[width=2 cm]{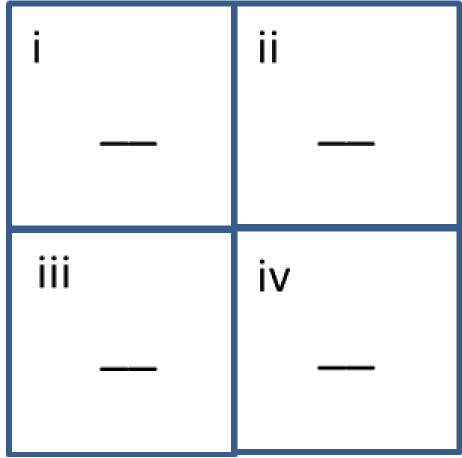}}}%
\label{b.12}%
\end{equation}
where $\phi_{\mathrm{i}}(\_),\;\phi_{\mathrm{ii}}(\_)\ldots$ denote the orbitals considered as functions of arbitrary position arguments. To avoid possible confusion with used notation we enumerate the orbital functions by Roman numbers.

The cells of Young diagrams can be associated with various possible orbital sets even within finite orbital configuration. In the original Heitler-London model the non-orthogonal on-site orbitals were suggested such that their overlap had provided the exchange mechanism as a physical resource for the empiric valence bond theory. However the advanced self-consistent Hartree-Fock method operates with the set of orthogonal orbitals, which we further assume.

By making use of the Young symmetrizers for a concrete orbital set, as indicated in the diagram cells of (\ref{b.11}) and (\ref{b.12}), we arrive at the following orbital compositions for the ansatz position wavefunctions
\begin{widetext}
\begin{eqnarray}
\Phi^{(0;1/2)}(i;j,k)&\propto& \vcenter{\hbox{\includegraphics[width=2.8 cm]{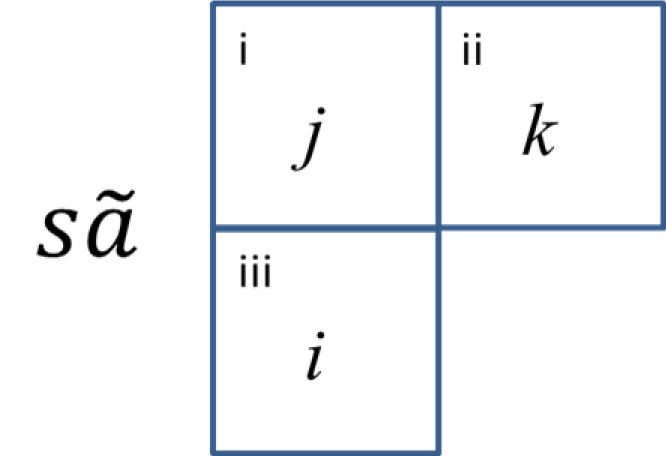}}}
=\left[\phi_{\mathrm{i}}(j)\,\phi_{\mathrm{iii}}(i)-\phi_{\mathrm{i}}(i)\,\phi_{\mathrm{iii}}(j)\right]\,\phi_{\mathrm{ii}}(k)%
+ \left[\phi_{\mathrm{i}}(k)\,\phi_{\mathrm{iii}}(i)-\phi_{\mathrm{i}}(i)\,\phi_{\mathrm{iii}}(k)\right]\,\phi_{\mathrm{ii}}(j)%
\nonumber\\%
\nonumber\\%
\Phi^{(1;1/2)}(i;j,k)&\propto& \vcenter{\hbox{\includegraphics[width=2.8 cm]{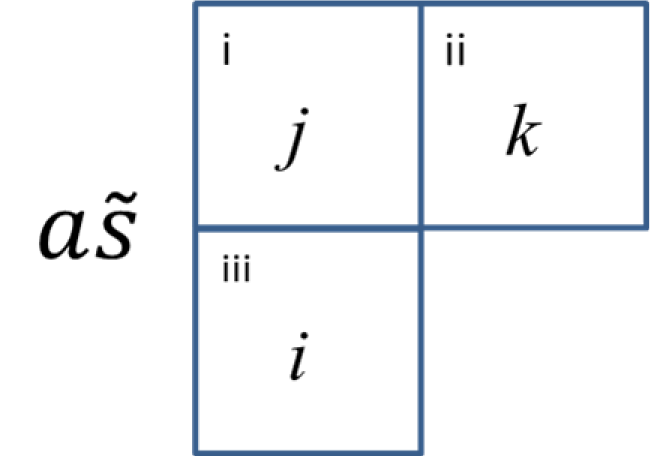}}}
=\left[\phi_{\mathrm{i}}(j)\,\phi_{\mathrm{iii}}(i)+\phi_{\mathrm{i}}(i)\,\phi_{\mathrm{iii}}(j)\right]\,\phi_{\mathrm{ii}}(k)%
- \left[\phi_{\mathrm{i}}(k)\,\phi_{\mathrm{iii}}(i)+\phi_{\mathrm{i}}(i)\,\phi_{\mathrm{iii}}(k)\right]\,\phi_{\mathrm{ii}}(j)%
\nonumber\\
\label{b.13}%
\end{eqnarray}
and
\begin{eqnarray}
\lefteqn{\Phi^{(0,0;0)}(i,j;k,l)\propto\vcenter{\hbox{\includegraphics[width=2.8 cm]{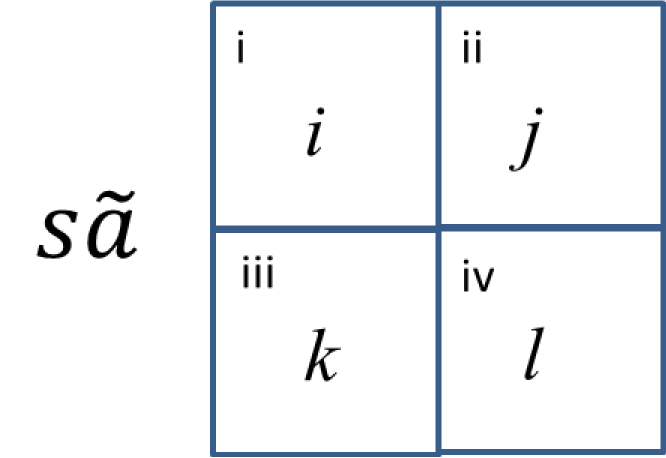}}}=}
\nonumber\\%
&& +\left[\phi_{\mathrm{i}}(i)\phi_{\mathrm{iii}}(k)-\phi_{\mathrm{i}}(k)\phi_{\mathrm{iii}}(i)\right]\left[\phi_{\mathrm{ii}}(j)\phi_{\mathrm{iv}}(l)-\phi_{\mathrm{ii}}(l)\phi_{\mathrm{iv}}(j)\right]%
+ \left[\phi_{\mathrm{i}}(j)\phi_{\mathrm{iii}}(k)-\phi_{\mathrm{i}}(k)\phi_{\mathrm{iii}}(j)\right]\left[\phi_{\mathrm{ii}}(i)\phi_{\mathrm{iv}}(l)-\phi_{\mathrm{ii}}(l)\phi_{\mathrm{iv}}(i)\right]%
\nonumber\\%
&&+ \left[\phi_{\mathrm{i}}(i)\phi_{\mathrm{iii}}(l)-\phi_{\mathrm{i}}(l)\phi_{\mathrm{iii}}(i)\right]\left[\phi_{\mathrm{ii}}(j)\phi_{\mathrm{iv}}(k)-\phi_{\mathrm{ii}}(k)\phi_{\mathrm{iv}}(j)\right]%
+ \left[\phi_{\mathrm{i}}(j)\phi_{\mathrm{iii}}(l)-\phi_{\mathrm{i}}(l)\phi_{\mathrm{iii}}(j)\right]\left[\phi_{\mathrm{ii}}(i)\phi_{\mathrm{iv}}(k)-\phi_{\mathrm{ii}}(k)\phi_{\mathrm{iv}}(i)\right]%
\nonumber\\%
\nonumber\\%
\lefteqn{\Phi^{(1,1;0)}(i,j;k,l)\propto\vcenter{\hbox{\includegraphics[width=2.8 cm]{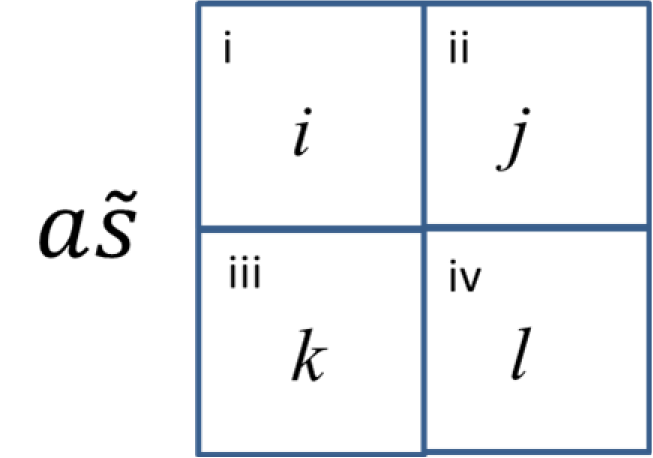}}}=}
\nonumber\\%
&& +\left[\phi_{\mathrm{i}}(i)\phi_{\mathrm{iii}}(k)+\phi_{\mathrm{i}}(k)\phi_{\mathrm{iii}}(i)\right]\left[\phi_{\mathrm{ii}}(j)\phi_{\mathrm{iv}}(l)+\phi_{\mathrm{ii}}(l)\phi_{\mathrm{iv}}(j)\right]%
- \left[\phi_{\mathrm{i}}(j)\phi_{\mathrm{iii}}(k)+\phi_{\mathrm{i}}(k)\phi_{\mathrm{iii}}(j)\right]\left[\phi_{\mathrm{ii}}(i)\phi_{\mathrm{iv}}(l)+\phi_{\mathrm{ii}}(l)\phi_{\mathrm{iv}}(i)\right]%
\nonumber\\%
&& - \left[\phi_{\mathrm{i}}(i)\phi_{\mathrm{iii}}(l)+\phi_{\mathrm{i}}(l)\phi_{\mathrm{iii}}(i)\right]\left[\phi_{\mathrm{ii}}(j)\phi_{\mathrm{iv}}(k)+\phi_{\mathrm{ii}}(k)\phi_{\mathrm{iv}}(j)\right]%
+ \left[\phi_{\mathrm{i}}(j)\phi_{\mathrm{iii}}(l)+\phi_{\mathrm{i}}(l)\phi_{\mathrm{iii}}(j)\right]\left[\phi_{\mathrm{ii}}(i)\phi_{\mathrm{iv}}(k)+\phi_{\mathrm{ii}}(k)\phi_{\mathrm{iv}}(i)\right]%
\nonumber\\%
\label{b.14}%
\end{eqnarray}
\end{widetext}
where "$s$" and "$a$" are respectively symmetrizer and anti-symmetrizer acting on rows (if no tilde) or columns (if tilde). We have skipped the normalization factor and ordered the operator actions in (\ref{b.13}) and (\ref{b.14}) in accordance with the symmetrization rules coordinated with construction of the conjugated spin wavefunctions, see above. 

As we can see, the defined position wavefunctions are critically sensitive to regrouping the orbitals within an accessible configuration i.e. to interchange $\phi_{\mathrm{i}}(\_)\leftrightharpoons\phi_{\mathrm{ii}}(\_)$, $\phi_{\mathrm{i}}(\_)\leftrightharpoons\phi_{\mathrm{iii}}(\_)$ etc. An optimal ansatz function for the ground state should contain the doubly occupied low energy orbital and either singly or doubly occupied nearest excited orbital. In that case $\phi_{\mathrm{i}}(\_)=\phi_{\mathrm{ii}}(\_)\equiv\phi_{g}(\_)$ and $\phi_{\mathrm{iii}}(\_)=\phi_{\mathrm{iv}}(\_)\equiv\phi_{e}(\_)$ would provide such optimum. 

However the system dynamics is generally reproducible by a wave-packet superposed in many such states with involving many orbital configurations and critically depends on taken reaction paths. That makes quite difficult its simulation by any classically served algorithms and computation protocols. The proposed quantum simulator can visualize and track important features of the process by relatively simple operating with microtraps and bosonic matter waves.

\bibliography{references}

\begin{thebibliography}{32}%
\makeatletter
\providecommand \@ifxundefined [1]{%
 \@ifx{#1\undefined}
}%
\providecommand \@ifnum [1]{%
 \ifnum #1\expandafter \@firstoftwo
 \else \expandafter \@secondoftwo
 \fi
}%
\providecommand \@ifx [1]{%
 \ifx #1\expandafter \@firstoftwo
 \else \expandafter \@secondoftwo
 \fi
}%
\providecommand \natexlab [1]{#1}%
\providecommand \enquote  [1]{``#1''}%
\providecommand \bibnamefont  [1]{#1}%
\providecommand \bibfnamefont [1]{#1}%
\providecommand \citenamefont [1]{#1}%
\providecommand \href@noop [0]{\@secondoftwo}%
\providecommand \href [0]{\begingroup \@sanitize@url \@href}%
\providecommand \@href[1]{\@@startlink{#1}\@@href}%
\providecommand \@@href[1]{\endgroup#1\@@endlink}%
\providecommand \@sanitize@url [0]{\catcode `\\12\catcode `\$12\catcode `\&12\catcode `\#12\catcode `\^12\catcode `\_12\catcode `\%12\relax}%
\providecommand \@@startlink[1]{}%
\providecommand \@@endlink[0]{}%
\providecommand \url  [0]{\begingroup\@sanitize@url \@url }%
\providecommand \@url [1]{\endgroup\@href {#1}{\urlprefix }}%
\providecommand \urlprefix  [0]{URL }%
\providecommand \Eprint [0]{\href }%
\providecommand \doibase [0]{https://doi.org/}%
\providecommand \selectlanguage [0]{\@gobble}%
\providecommand \bibinfo  [0]{\@secondoftwo}%
\providecommand \bibfield  [0]{\@secondoftwo}%
\providecommand \translation [1]{[#1]}%
\providecommand \BibitemOpen [0]{}%
\providecommand \bibitemStop [0]{}%
\providecommand \bibitemNoStop [0]{.\EOS\space}%
\providecommand \EOS [0]{\spacefactor3000\relax}%
\providecommand \BibitemShut  [1]{\csname bibitem#1\endcsname}%
\let\auto@bib@innerbib\@empty
\bibitem [{\citenamefont {Bernien}\ \emph {et~al.}(2017)\citenamefont {Bernien}, \citenamefont {Schwartz}, \citenamefont {Keesling}, \citenamefont {Levine}, \citenamefont {Omran}, \citenamefont {Pichler}, \citenamefont {Choi}, \citenamefont {Zibrov}, \citenamefont {Endres}, \citenamefont {Greiner}, \citenamefont {Vuletić},\ and\ \citenamefont {Lukin}}]{Lukin2017}%
  \BibitemOpen
  \bibfield  {author} {\bibinfo {author} {\bibfnamefont {H.}~\bibnamefont {Bernien}}, \bibinfo {author} {\bibfnamefont {S.}~\bibnamefont {Schwartz}}, \bibinfo {author} {\bibfnamefont {A.}~\bibnamefont {Keesling}}, \bibinfo {author} {\bibfnamefont {H.}~\bibnamefont {Levine}}, \bibinfo {author} {\bibfnamefont {A.}~\bibnamefont {Omran}}, \bibinfo {author} {\bibfnamefont {H.}~\bibnamefont {Pichler}}, \bibinfo {author} {\bibfnamefont {S.}~\bibnamefont {Choi}}, \bibinfo {author} {\bibfnamefont {A.~S.}\ \bibnamefont {Zibrov}}, \bibinfo {author} {\bibfnamefont {M.}~\bibnamefont {Endres}}, \bibinfo {author} {\bibfnamefont {M.}~\bibnamefont {Greiner}}, \bibinfo {author} {\bibfnamefont {V.}~\bibnamefont {Vuletić}},\ and\ \bibinfo {author} {\bibfnamefont {M.~D.}\ \bibnamefont {Lukin}},\ }\bibfield  {title} {\bibinfo {title} {Probing many-body dynamics on a 51-atom quantum simulator},\ }\href {https://doi.org/10.1038/nature24622} {\bibfield  {journal} {\bibinfo  {journal} {Nature}\ }\textbf {\bibinfo {volume} {551}},\
  \bibinfo {pages} {579–584} (\bibinfo {year} {2017})}\BibitemShut {NoStop}%
\bibitem [{\citenamefont {Ebadi}\ \emph {et~al.}(2021)\citenamefont {Ebadi}, \citenamefont {Wang}, \citenamefont {Levine}, \citenamefont {Keesling}, \citenamefont {Semeghini}, \citenamefont {Omran}, \citenamefont {Bluvstein}, \citenamefont {Samajdar}, \citenamefont {Pichler}, \citenamefont {Ho}, \citenamefont {Choi}, \citenamefont {Sachdev}, \citenamefont {Greiner}, \citenamefont {Vuletić},\ and\ \citenamefont {Lukin}}]{Lukin2021}%
  \BibitemOpen
  \bibfield  {author} {\bibinfo {author} {\bibfnamefont {S.}~\bibnamefont {Ebadi}}, \bibinfo {author} {\bibfnamefont {T.~T.}\ \bibnamefont {Wang}}, \bibinfo {author} {\bibfnamefont {H.}~\bibnamefont {Levine}}, \bibinfo {author} {\bibfnamefont {A.}~\bibnamefont {Keesling}}, \bibinfo {author} {\bibfnamefont {G.}~\bibnamefont {Semeghini}}, \bibinfo {author} {\bibfnamefont {A.}~\bibnamefont {Omran}}, \bibinfo {author} {\bibfnamefont {D.}~\bibnamefont {Bluvstein}}, \bibinfo {author} {\bibfnamefont {R.}~\bibnamefont {Samajdar}}, \bibinfo {author} {\bibfnamefont {H.}~\bibnamefont {Pichler}}, \bibinfo {author} {\bibfnamefont {W.~W.}\ \bibnamefont {Ho}}, \bibinfo {author} {\bibfnamefont {S.}~\bibnamefont {Choi}}, \bibinfo {author} {\bibfnamefont {S.}~\bibnamefont {Sachdev}}, \bibinfo {author} {\bibfnamefont {M.}~\bibnamefont {Greiner}}, \bibinfo {author} {\bibfnamefont {V.}~\bibnamefont {Vuletić}},\ and\ \bibinfo {author} {\bibfnamefont {M.~D.}\ \bibnamefont {Lukin}},\ }\bibfield  {title} {\bibinfo {title} {Quantum
  phases of matter on a 256-atom programmable quantum simulator},\ }\href {https://doi.org/10.1038/s41586-021-03582-4} {\bibfield  {journal} {\bibinfo  {journal} {Nature}\ }\textbf {\bibinfo {volume} {595}},\ \bibinfo {pages} {227–232} (\bibinfo {year} {2021})}\BibitemShut {NoStop}%
\bibitem [{\citenamefont {Keesling}\ \emph {et~al.}(2019)\citenamefont {Keesling}, \citenamefont {Omran}, \citenamefont {Levine}, \citenamefont {Bernien}, \citenamefont {Pichler}, \citenamefont {Choi}, \citenamefont {Samajdar}, \citenamefont {Schwartz}, \citenamefont {Silvi}, \citenamefont {Sachdev}, \citenamefont {Zoller}, \citenamefont {Endres}, \citenamefont {Greiner}, \citenamefont {Vuletić},\ and\ \citenamefont {Lukin}}]{Lukin2019}%
  \BibitemOpen
  \bibfield  {author} {\bibinfo {author} {\bibfnamefont {A.}~\bibnamefont {Keesling}}, \bibinfo {author} {\bibfnamefont {A.}~\bibnamefont {Omran}}, \bibinfo {author} {\bibfnamefont {H.}~\bibnamefont {Levine}}, \bibinfo {author} {\bibfnamefont {H.}~\bibnamefont {Bernien}}, \bibinfo {author} {\bibfnamefont {H.}~\bibnamefont {Pichler}}, \bibinfo {author} {\bibfnamefont {S.}~\bibnamefont {Choi}}, \bibinfo {author} {\bibfnamefont {R.}~\bibnamefont {Samajdar}}, \bibinfo {author} {\bibfnamefont {S.}~\bibnamefont {Schwartz}}, \bibinfo {author} {\bibfnamefont {P.}~\bibnamefont {Silvi}}, \bibinfo {author} {\bibfnamefont {S.}~\bibnamefont {Sachdev}}, \bibinfo {author} {\bibfnamefont {P.}~\bibnamefont {Zoller}}, \bibinfo {author} {\bibfnamefont {M.}~\bibnamefont {Endres}}, \bibinfo {author} {\bibfnamefont {M.}~\bibnamefont {Greiner}}, \bibinfo {author} {\bibfnamefont {V.}~\bibnamefont {Vuletić}},\ and\ \bibinfo {author} {\bibfnamefont {M.~D.}\ \bibnamefont {Lukin}},\ }\bibfield  {title} {\bibinfo {title} {Quantum
  kibble–zurek mechanism and critical dynamics on a programmable rydberg simulator},\ }\href {https://doi.org/10.1038/s41586-019-1070-1} {\bibfield  {journal} {\bibinfo  {journal} {Nature}\ }\textbf {\bibinfo {volume} {568}},\ \bibinfo {pages} {207–211} (\bibinfo {year} {2019})}\BibitemShut {NoStop}%
\bibitem [{\citenamefont {Bluvstein}\ \emph {et~al.}(2022)\citenamefont {Bluvstein}, \citenamefont {Levine}, \citenamefont {Semeghini}, \citenamefont {Wang}, \citenamefont {Ebadi}, \citenamefont {Kalinowski}, \citenamefont {Keesling}, \citenamefont {Maskara}, \citenamefont {Pichler}, \citenamefont {Greiner}, \citenamefont {Vuletić},\ and\ \citenamefont {Lukin}}]{Lukin2022}%
  \BibitemOpen
  \bibfield  {author} {\bibinfo {author} {\bibfnamefont {D.}~\bibnamefont {Bluvstein}}, \bibinfo {author} {\bibfnamefont {H.}~\bibnamefont {Levine}}, \bibinfo {author} {\bibfnamefont {G.}~\bibnamefont {Semeghini}}, \bibinfo {author} {\bibfnamefont {T.~T.}\ \bibnamefont {Wang}}, \bibinfo {author} {\bibfnamefont {S.}~\bibnamefont {Ebadi}}, \bibinfo {author} {\bibfnamefont {M.}~\bibnamefont {Kalinowski}}, \bibinfo {author} {\bibfnamefont {A.}~\bibnamefont {Keesling}}, \bibinfo {author} {\bibfnamefont {N.}~\bibnamefont {Maskara}}, \bibinfo {author} {\bibfnamefont {H.}~\bibnamefont {Pichler}}, \bibinfo {author} {\bibfnamefont {M.}~\bibnamefont {Greiner}}, \bibinfo {author} {\bibfnamefont {V.}~\bibnamefont {Vuletić}},\ and\ \bibinfo {author} {\bibfnamefont {M.~D.}\ \bibnamefont {Lukin}},\ }\bibfield  {title} {\bibinfo {title} {A quantum processor based on coherent transport of entangled atom arrays},\ }\href {https://doi.org/10.1038/s41586-022-04592-6} {\bibfield  {journal} {\bibinfo  {journal} {Nature}\ }\textbf
  {\bibinfo {volume} {604}},\ \bibinfo {pages} {451–456} (\bibinfo {year} {2022})}\BibitemShut {NoStop}%
\bibitem [{\citenamefont {Radnaev}\ \emph {et~al.}(2024)\citenamefont {Radnaev}, \citenamefont {Chung}, \citenamefont {Cole}, \citenamefont {Mason}, \citenamefont {Ballance}, \citenamefont {Bedalov}, \citenamefont {Belknap}, \citenamefont {Berman}, \citenamefont {Blakely}, \citenamefont {Bloomfield}, \citenamefont {Buttler}, \citenamefont {Campbell}, \citenamefont {Chopinaud}, \citenamefont {Copenhaver}, \citenamefont {Dawes}, \citenamefont {Eubanks}, \citenamefont {Friss}, \citenamefont {Garcia}, \citenamefont {Gilbert}, \citenamefont {Gillette}, \citenamefont {Goiporia}, \citenamefont {Gokhale}, \citenamefont {Goldwin}, \citenamefont {Goodwin}, \citenamefont {Graham}, \citenamefont {Guttormsson}, \citenamefont {Hickman}, \citenamefont {Hurtley}, \citenamefont {Iliev}, \citenamefont {Jones}, \citenamefont {Jones}, \citenamefont {Kuper}, \citenamefont {Lewis}, \citenamefont {Lichtman}, \citenamefont {Majdeteimouri}, \citenamefont {Mason}, \citenamefont {McMaster}, \citenamefont {Miles}, \citenamefont
  {Mitchell}, \citenamefont {Murphree}, \citenamefont {Neff-Mallon}, \citenamefont {Oh}, \citenamefont {Omole}, \citenamefont {Simon}, \citenamefont {Pederson}, \citenamefont {Perlin}, \citenamefont {Reiter}, \citenamefont {Rines}, \citenamefont {Romlow}, \citenamefont {Scott}, \citenamefont {Stiefvater}, \citenamefont {Tanner}, \citenamefont {Tucker}, \citenamefont {Vinogradov}, \citenamefont {Warter}, \citenamefont {Yeo}, \citenamefont {Saffman},\ and\ \citenamefont {Noel}}]{Radnaev2024}%
  \BibitemOpen
  \bibfield  {author} {\bibinfo {author} {\bibfnamefont {A.~G.}\ \bibnamefont {Radnaev}}, \bibinfo {author} {\bibfnamefont {W.~C.}\ \bibnamefont {Chung}}, \bibinfo {author} {\bibfnamefont {D.~C.}\ \bibnamefont {Cole}}, \bibinfo {author} {\bibfnamefont {D.}~\bibnamefont {Mason}}, \bibinfo {author} {\bibfnamefont {T.~G.}\ \bibnamefont {Ballance}}, \bibinfo {author} {\bibfnamefont {M.~J.}\ \bibnamefont {Bedalov}}, \bibinfo {author} {\bibfnamefont {D.~A.}\ \bibnamefont {Belknap}}, \bibinfo {author} {\bibfnamefont {M.~R.}\ \bibnamefont {Berman}}, \bibinfo {author} {\bibfnamefont {M.}~\bibnamefont {Blakely}}, \bibinfo {author} {\bibfnamefont {I.~L.}\ \bibnamefont {Bloomfield}}, \bibinfo {author} {\bibfnamefont {P.~D.}\ \bibnamefont {Buttler}}, \bibinfo {author} {\bibfnamefont {C.}~\bibnamefont {Campbell}}, \bibinfo {author} {\bibfnamefont {A.}~\bibnamefont {Chopinaud}}, \bibinfo {author} {\bibfnamefont {E.}~\bibnamefont {Copenhaver}}, \bibinfo {author} {\bibfnamefont {M.~K.}\ \bibnamefont {Dawes}}, \bibinfo
  {author} {\bibfnamefont {S.~Y.}\ \bibnamefont {Eubanks}}, \bibinfo {author} {\bibfnamefont {A.~J.}\ \bibnamefont {Friss}}, \bibinfo {author} {\bibfnamefont {D.~M.}\ \bibnamefont {Garcia}}, \bibinfo {author} {\bibfnamefont {J.}~\bibnamefont {Gilbert}}, \bibinfo {author} {\bibfnamefont {M.}~\bibnamefont {Gillette}}, \bibinfo {author} {\bibfnamefont {P.}~\bibnamefont {Goiporia}}, \bibinfo {author} {\bibfnamefont {P.}~\bibnamefont {Gokhale}}, \bibinfo {author} {\bibfnamefont {J.}~\bibnamefont {Goldwin}}, \bibinfo {author} {\bibfnamefont {D.}~\bibnamefont {Goodwin}}, \bibinfo {author} {\bibfnamefont {T.~M.}\ \bibnamefont {Graham}}, \bibinfo {author} {\bibfnamefont {C.}~\bibnamefont {Guttormsson}}, \bibinfo {author} {\bibfnamefont {G.~T.}\ \bibnamefont {Hickman}}, \bibinfo {author} {\bibfnamefont {L.}~\bibnamefont {Hurtley}}, \bibinfo {author} {\bibfnamefont {M.}~\bibnamefont {Iliev}}, \bibinfo {author} {\bibfnamefont {E.~B.}\ \bibnamefont {Jones}}, \bibinfo {author} {\bibfnamefont {R.~A.}\ \bibnamefont {Jones}},
  \bibinfo {author} {\bibfnamefont {K.~W.}\ \bibnamefont {Kuper}}, \bibinfo {author} {\bibfnamefont {T.~B.}\ \bibnamefont {Lewis}}, \bibinfo {author} {\bibfnamefont {M.~T.}\ \bibnamefont {Lichtman}}, \bibinfo {author} {\bibfnamefont {F.}~\bibnamefont {Majdeteimouri}}, \bibinfo {author} {\bibfnamefont {J.~J.}\ \bibnamefont {Mason}}, \bibinfo {author} {\bibfnamefont {J.~K.}\ \bibnamefont {McMaster}}, \bibinfo {author} {\bibfnamefont {J.~A.}\ \bibnamefont {Miles}}, \bibinfo {author} {\bibfnamefont {P.~T.}\ \bibnamefont {Mitchell}}, \bibinfo {author} {\bibfnamefont {J.~D.}\ \bibnamefont {Murphree}}, \bibinfo {author} {\bibfnamefont {N.~A.}\ \bibnamefont {Neff-Mallon}}, \bibinfo {author} {\bibfnamefont {T.}~\bibnamefont {Oh}}, \bibinfo {author} {\bibfnamefont {V.}~\bibnamefont {Omole}}, \bibinfo {author} {\bibfnamefont {C.~P.}\ \bibnamefont {Simon}}, \bibinfo {author} {\bibfnamefont {N.}~\bibnamefont {Pederson}}, \bibinfo {author} {\bibfnamefont {M.~A.}\ \bibnamefont {Perlin}}, \bibinfo {author} {\bibfnamefont
  {A.}~\bibnamefont {Reiter}}, \bibinfo {author} {\bibfnamefont {R.}~\bibnamefont {Rines}}, \bibinfo {author} {\bibfnamefont {P.}~\bibnamefont {Romlow}}, \bibinfo {author} {\bibfnamefont {A.~M.}\ \bibnamefont {Scott}}, \bibinfo {author} {\bibfnamefont {D.}~\bibnamefont {Stiefvater}}, \bibinfo {author} {\bibfnamefont {J.~R.}\ \bibnamefont {Tanner}}, \bibinfo {author} {\bibfnamefont {A.~K.}\ \bibnamefont {Tucker}}, \bibinfo {author} {\bibfnamefont {I.~V.}\ \bibnamefont {Vinogradov}}, \bibinfo {author} {\bibfnamefont {M.~L.}\ \bibnamefont {Warter}}, \bibinfo {author} {\bibfnamefont {M.}~\bibnamefont {Yeo}}, \bibinfo {author} {\bibfnamefont {M.}~\bibnamefont {Saffman}},\ and\ \bibinfo {author} {\bibfnamefont {T.~W.}\ \bibnamefont {Noel}},\ }\href {https://arxiv.org/abs/2408.08288} {\bibinfo {title} {A universal neutral-atom quantum computer with individual optical addressing and non-destructive readout}} (\bibinfo {year} {2024}),\ \Eprint {https://arxiv.org/abs/2408.08288} {arXiv:2408.08288 [quant-ph]}
  \BibitemShut {NoStop}%
\bibitem [{\citenamefont {Reichardt}\ \emph {et~al.}(2024)\citenamefont {Reichardt}, \citenamefont {Paetznick}, \citenamefont {Aasen}, \citenamefont {Basov}, \citenamefont {Bello-Rivas}, \citenamefont {Bonderson}, \citenamefont {Chao}, \citenamefont {van Dam}, \citenamefont {Hastings}, \citenamefont {Paz}, \citenamefont {da~Silva}, \citenamefont {Sundaram}, \citenamefont {Svore}, \citenamefont {Vaschillo}, \citenamefont {Wang}, \citenamefont {Zanner}, \citenamefont {Cairncross}, \citenamefont {Chen}, \citenamefont {Crow}, \citenamefont {Kim}, \citenamefont {Kindem}, \citenamefont {King}, \citenamefont {McDonald}, \citenamefont {Norcia}, \citenamefont {Ryou}, \citenamefont {Stone}, \citenamefont {Wadleigh}, \citenamefont {Barnes}, \citenamefont {Battaglino}, \citenamefont {Bohdanowicz}, \citenamefont {Booth}, \citenamefont {Brown}, \citenamefont {Brown}, \citenamefont {Cassella}, \citenamefont {Coxe}, \citenamefont {Epstein}, \citenamefont {Feldkamp}, \citenamefont {Griger}, \citenamefont {Halperin},
  \citenamefont {Heinz}, \citenamefont {Hummel}, \citenamefont {Jaffe}, \citenamefont {Jones}, \citenamefont {Kapit}, \citenamefont {Kotru}, \citenamefont {Lauigan}, \citenamefont {Li}, \citenamefont {Marjanovic}, \citenamefont {Megidish}, \citenamefont {Meredith}, \citenamefont {Morshead}, \citenamefont {Muniz}, \citenamefont {Narayanaswami}, \citenamefont {Nishiguchi}, \citenamefont {Paule}, \citenamefont {Pawlak}, \citenamefont {Pudenz}, \citenamefont {Pérez}, \citenamefont {Simon}, \citenamefont {Smull}, \citenamefont {Stack}, \citenamefont {Urbanek}, \citenamefont {van~de Veerdonk}, \citenamefont {Vendeiro}, \citenamefont {Weverka}, \citenamefont {Wilkason}, \citenamefont {Wu}, \citenamefont {Xie}, \citenamefont {Zalys-Geller}, \citenamefont {Zhang},\ and\ \citenamefont {Bloom}}]{Reichardt2024}%
  \BibitemOpen
  \bibfield  {author} {\bibinfo {author} {\bibfnamefont {B.~W.}\ \bibnamefont {Reichardt}}, \bibinfo {author} {\bibfnamefont {A.}~\bibnamefont {Paetznick}}, \bibinfo {author} {\bibfnamefont {D.}~\bibnamefont {Aasen}}, \bibinfo {author} {\bibfnamefont {I.}~\bibnamefont {Basov}}, \bibinfo {author} {\bibfnamefont {J.~M.}\ \bibnamefont {Bello-Rivas}}, \bibinfo {author} {\bibfnamefont {P.}~\bibnamefont {Bonderson}}, \bibinfo {author} {\bibfnamefont {R.}~\bibnamefont {Chao}}, \bibinfo {author} {\bibfnamefont {W.}~\bibnamefont {van Dam}}, \bibinfo {author} {\bibfnamefont {M.~B.}\ \bibnamefont {Hastings}}, \bibinfo {author} {\bibfnamefont {A.}~\bibnamefont {Paz}}, \bibinfo {author} {\bibfnamefont {M.~P.}\ \bibnamefont {da~Silva}}, \bibinfo {author} {\bibfnamefont {A.}~\bibnamefont {Sundaram}}, \bibinfo {author} {\bibfnamefont {K.~M.}\ \bibnamefont {Svore}}, \bibinfo {author} {\bibfnamefont {A.}~\bibnamefont {Vaschillo}}, \bibinfo {author} {\bibfnamefont {Z.}~\bibnamefont {Wang}}, \bibinfo {author} {\bibfnamefont
  {M.}~\bibnamefont {Zanner}}, \bibinfo {author} {\bibfnamefont {W.~B.}\ \bibnamefont {Cairncross}}, \bibinfo {author} {\bibfnamefont {C.-A.}\ \bibnamefont {Chen}}, \bibinfo {author} {\bibfnamefont {D.}~\bibnamefont {Crow}}, \bibinfo {author} {\bibfnamefont {H.}~\bibnamefont {Kim}}, \bibinfo {author} {\bibfnamefont {J.~M.}\ \bibnamefont {Kindem}}, \bibinfo {author} {\bibfnamefont {J.}~\bibnamefont {King}}, \bibinfo {author} {\bibfnamefont {M.}~\bibnamefont {McDonald}}, \bibinfo {author} {\bibfnamefont {M.~A.}\ \bibnamefont {Norcia}}, \bibinfo {author} {\bibfnamefont {A.}~\bibnamefont {Ryou}}, \bibinfo {author} {\bibfnamefont {M.}~\bibnamefont {Stone}}, \bibinfo {author} {\bibfnamefont {L.}~\bibnamefont {Wadleigh}}, \bibinfo {author} {\bibfnamefont {K.}~\bibnamefont {Barnes}}, \bibinfo {author} {\bibfnamefont {P.}~\bibnamefont {Battaglino}}, \bibinfo {author} {\bibfnamefont {T.~C.}\ \bibnamefont {Bohdanowicz}}, \bibinfo {author} {\bibfnamefont {G.}~\bibnamefont {Booth}}, \bibinfo {author} {\bibfnamefont
  {A.}~\bibnamefont {Brown}}, \bibinfo {author} {\bibfnamefont {M.~O.}\ \bibnamefont {Brown}}, \bibinfo {author} {\bibfnamefont {K.}~\bibnamefont {Cassella}}, \bibinfo {author} {\bibfnamefont {R.}~\bibnamefont {Coxe}}, \bibinfo {author} {\bibfnamefont {J.~M.}\ \bibnamefont {Epstein}}, \bibinfo {author} {\bibfnamefont {M.}~\bibnamefont {Feldkamp}}, \bibinfo {author} {\bibfnamefont {C.}~\bibnamefont {Griger}}, \bibinfo {author} {\bibfnamefont {E.}~\bibnamefont {Halperin}}, \bibinfo {author} {\bibfnamefont {A.}~\bibnamefont {Heinz}}, \bibinfo {author} {\bibfnamefont {F.}~\bibnamefont {Hummel}}, \bibinfo {author} {\bibfnamefont {M.}~\bibnamefont {Jaffe}}, \bibinfo {author} {\bibfnamefont {A.~M.~W.}\ \bibnamefont {Jones}}, \bibinfo {author} {\bibfnamefont {E.}~\bibnamefont {Kapit}}, \bibinfo {author} {\bibfnamefont {K.}~\bibnamefont {Kotru}}, \bibinfo {author} {\bibfnamefont {J.}~\bibnamefont {Lauigan}}, \bibinfo {author} {\bibfnamefont {M.}~\bibnamefont {Li}}, \bibinfo {author} {\bibfnamefont {J.}~\bibnamefont
  {Marjanovic}}, \bibinfo {author} {\bibfnamefont {E.}~\bibnamefont {Megidish}}, \bibinfo {author} {\bibfnamefont {M.}~\bibnamefont {Meredith}}, \bibinfo {author} {\bibfnamefont {R.}~\bibnamefont {Morshead}}, \bibinfo {author} {\bibfnamefont {J.~A.}\ \bibnamefont {Muniz}}, \bibinfo {author} {\bibfnamefont {S.}~\bibnamefont {Narayanaswami}}, \bibinfo {author} {\bibfnamefont {C.}~\bibnamefont {Nishiguchi}}, \bibinfo {author} {\bibfnamefont {T.}~\bibnamefont {Paule}}, \bibinfo {author} {\bibfnamefont {K.~A.}\ \bibnamefont {Pawlak}}, \bibinfo {author} {\bibfnamefont {K.~L.}\ \bibnamefont {Pudenz}}, \bibinfo {author} {\bibfnamefont {D.~R.}\ \bibnamefont {Pérez}}, \bibinfo {author} {\bibfnamefont {J.}~\bibnamefont {Simon}}, \bibinfo {author} {\bibfnamefont {A.}~\bibnamefont {Smull}}, \bibinfo {author} {\bibfnamefont {D.}~\bibnamefont {Stack}}, \bibinfo {author} {\bibfnamefont {M.}~\bibnamefont {Urbanek}}, \bibinfo {author} {\bibfnamefont {R.~J.~M.}\ \bibnamefont {van~de Veerdonk}}, \bibinfo {author} {\bibfnamefont
  {Z.}~\bibnamefont {Vendeiro}}, \bibinfo {author} {\bibfnamefont {R.~T.}\ \bibnamefont {Weverka}}, \bibinfo {author} {\bibfnamefont {T.}~\bibnamefont {Wilkason}}, \bibinfo {author} {\bibfnamefont {T.-Y.}\ \bibnamefont {Wu}}, \bibinfo {author} {\bibfnamefont {X.}~\bibnamefont {Xie}}, \bibinfo {author} {\bibfnamefont {E.}~\bibnamefont {Zalys-Geller}}, \bibinfo {author} {\bibfnamefont {X.}~\bibnamefont {Zhang}},\ and\ \bibinfo {author} {\bibfnamefont {B.~J.}\ \bibnamefont {Bloom}},\ }\href {https://arxiv.org/abs/2411.11822} {\bibinfo {title} {Logical computation demonstrated with a neutral atom quantum processor}} (\bibinfo {year} {2024}),\ \Eprint {https://arxiv.org/abs/2411.11822} {arXiv:2411.11822 [quant-ph]} \BibitemShut {NoStop}%
\bibitem [{\citenamefont {Cao}\ \emph {et~al.}(2019)\citenamefont {Cao}, \citenamefont {Romero}, \citenamefont {Olson}, \citenamefont {Degroote}, \citenamefont {Johnson}, \citenamefont {Kieferová}, \citenamefont {Kivlichan}, \citenamefont {Menke}, \citenamefont {Peropadre}, \citenamefont {Sawaya}, \citenamefont {Sim}, \citenamefont {Veis},\ and\ \citenamefont {Aspuru-Guzik}}]{Cao2019}%
  \BibitemOpen
  \bibfield  {author} {\bibinfo {author} {\bibfnamefont {Y.}~\bibnamefont {Cao}}, \bibinfo {author} {\bibfnamefont {J.}~\bibnamefont {Romero}}, \bibinfo {author} {\bibfnamefont {J.~P.}\ \bibnamefont {Olson}}, \bibinfo {author} {\bibfnamefont {M.}~\bibnamefont {Degroote}}, \bibinfo {author} {\bibfnamefont {P.~D.}\ \bibnamefont {Johnson}}, \bibinfo {author} {\bibfnamefont {M.}~\bibnamefont {Kieferová}}, \bibinfo {author} {\bibfnamefont {I.~D.}\ \bibnamefont {Kivlichan}}, \bibinfo {author} {\bibfnamefont {T.}~\bibnamefont {Menke}}, \bibinfo {author} {\bibfnamefont {B.}~\bibnamefont {Peropadre}}, \bibinfo {author} {\bibfnamefont {N.~P.~D.}\ \bibnamefont {Sawaya}}, \bibinfo {author} {\bibfnamefont {S.}~\bibnamefont {Sim}}, \bibinfo {author} {\bibfnamefont {L.}~\bibnamefont {Veis}},\ and\ \bibinfo {author} {\bibfnamefont {A.}~\bibnamefont {Aspuru-Guzik}},\ }\bibfield  {title} {\bibinfo {title} {Quantum chemistry in the age of quantum computing},\ }\href {https://doi.org/10.1021/acs.chemrev.8b00803} {\bibfield
  {journal} {\bibinfo  {journal} {Chemical Reviews}\ }\textbf {\bibinfo {volume} {119}},\ \bibinfo {pages} {10856–10915} (\bibinfo {year} {2019})}\BibitemShut {NoStop}%
\bibitem [{\citenamefont {Hempel}\ \emph {et~al.}(2018)\citenamefont {Hempel}, \citenamefont {Maier}, \citenamefont {Romero}, \citenamefont {McClean}, \citenamefont {Monz}, \citenamefont {Shen}, \citenamefont {Jurcevic}, \citenamefont {Lanyon}, \citenamefont {Love}, \citenamefont {Babbush}, \citenamefont {Aspuru-Guzik}, \citenamefont {Blatt},\ and\ \citenamefont {Roos}}]{Hempel2018}%
  \BibitemOpen
  \bibfield  {author} {\bibinfo {author} {\bibfnamefont {C.}~\bibnamefont {Hempel}}, \bibinfo {author} {\bibfnamefont {C.}~\bibnamefont {Maier}}, \bibinfo {author} {\bibfnamefont {J.}~\bibnamefont {Romero}}, \bibinfo {author} {\bibfnamefont {J.}~\bibnamefont {McClean}}, \bibinfo {author} {\bibfnamefont {T.}~\bibnamefont {Monz}}, \bibinfo {author} {\bibfnamefont {H.}~\bibnamefont {Shen}}, \bibinfo {author} {\bibfnamefont {P.}~\bibnamefont {Jurcevic}}, \bibinfo {author} {\bibfnamefont {B.~P.}\ \bibnamefont {Lanyon}}, \bibinfo {author} {\bibfnamefont {P.}~\bibnamefont {Love}}, \bibinfo {author} {\bibfnamefont {R.}~\bibnamefont {Babbush}}, \bibinfo {author} {\bibfnamefont {A.}~\bibnamefont {Aspuru-Guzik}}, \bibinfo {author} {\bibfnamefont {R.}~\bibnamefont {Blatt}},\ and\ \bibinfo {author} {\bibfnamefont {C.~F.}\ \bibnamefont {Roos}},\ }\bibfield  {title} {\bibinfo {title} {Quantum chemistry calculations on a trapped-ion quantum simulator},\ }\href {https://doi.org/10.1103/PhysRevX.8.031022} {\bibfield  {journal}
  {\bibinfo  {journal} {Phys. Rev. X}\ }\textbf {\bibinfo {volume} {8}},\ \bibinfo {pages} {031022} (\bibinfo {year} {2018})}\BibitemShut {NoStop}%
\bibitem [{\citenamefont {Motta}\ \emph {et~al.}(2023)\citenamefont {Motta}, \citenamefont {Jones}, \citenamefont {Rice}, \citenamefont {Gujarati}, \citenamefont {Sakuma}, \citenamefont {Liepuoniute}, \citenamefont {Garcia},\ and\ \citenamefont {Ohnishi}}]{Motta2023}%
  \BibitemOpen
  \bibfield  {author} {\bibinfo {author} {\bibfnamefont {M.}~\bibnamefont {Motta}}, \bibinfo {author} {\bibfnamefont {G.~O.}\ \bibnamefont {Jones}}, \bibinfo {author} {\bibfnamefont {J.~E.}\ \bibnamefont {Rice}}, \bibinfo {author} {\bibfnamefont {T.~P.}\ \bibnamefont {Gujarati}}, \bibinfo {author} {\bibfnamefont {R.}~\bibnamefont {Sakuma}}, \bibinfo {author} {\bibfnamefont {I.}~\bibnamefont {Liepuoniute}}, \bibinfo {author} {\bibfnamefont {J.~M.}\ \bibnamefont {Garcia}},\ and\ \bibinfo {author} {\bibfnamefont {Y.-y.}\ \bibnamefont {Ohnishi}},\ }\bibfield  {title} {\bibinfo {title} {Quantum chemistry simulation of ground- and excited-state properties of the sulfonium cation on a superconducting quantum processor},\ }\href {https://doi.org/10.1039/D2SC06019A} {\bibfield  {journal} {\bibinfo  {journal} {Chem. Sci.}\ }\textbf {\bibinfo {volume} {14}},\ \bibinfo {pages} {2915} (\bibinfo {year} {2023})}\BibitemShut {NoStop}%
\bibitem [{\citenamefont {Robledo-Moreno}\ \emph {et~al.}(2024)\citenamefont {Robledo-Moreno}, \citenamefont {Motta}, \citenamefont {Haas}, \citenamefont {Javadi-Abhari}, \citenamefont {Jurcevic}, \citenamefont {Kirby}, \citenamefont {Martiel}, \citenamefont {Sharma}, \citenamefont {Sharma}, \citenamefont {Shirakawa}, \citenamefont {Sitdikov}, \citenamefont {Sun}, \citenamefont {Sung}, \citenamefont {Takita}, \citenamefont {Tran}, \citenamefont {Yunoki},\ and\ \citenamefont {Mezzacapo}}]{Robledo-Moreno2024}%
  \BibitemOpen
  \bibfield  {author} {\bibinfo {author} {\bibfnamefont {J.}~\bibnamefont {Robledo-Moreno}}, \bibinfo {author} {\bibfnamefont {M.}~\bibnamefont {Motta}}, \bibinfo {author} {\bibfnamefont {H.}~\bibnamefont {Haas}}, \bibinfo {author} {\bibfnamefont {A.}~\bibnamefont {Javadi-Abhari}}, \bibinfo {author} {\bibfnamefont {P.}~\bibnamefont {Jurcevic}}, \bibinfo {author} {\bibfnamefont {W.}~\bibnamefont {Kirby}}, \bibinfo {author} {\bibfnamefont {S.}~\bibnamefont {Martiel}}, \bibinfo {author} {\bibfnamefont {K.}~\bibnamefont {Sharma}}, \bibinfo {author} {\bibfnamefont {S.}~\bibnamefont {Sharma}}, \bibinfo {author} {\bibfnamefont {T.}~\bibnamefont {Shirakawa}}, \bibinfo {author} {\bibfnamefont {I.}~\bibnamefont {Sitdikov}}, \bibinfo {author} {\bibfnamefont {R.-Y.}\ \bibnamefont {Sun}}, \bibinfo {author} {\bibfnamefont {K.~J.}\ \bibnamefont {Sung}}, \bibinfo {author} {\bibfnamefont {M.}~\bibnamefont {Takita}}, \bibinfo {author} {\bibfnamefont {M.~C.}\ \bibnamefont {Tran}}, \bibinfo {author} {\bibfnamefont
  {S.}~\bibnamefont {Yunoki}},\ and\ \bibinfo {author} {\bibfnamefont {A.}~\bibnamefont {Mezzacapo}},\ }\href {https://arxiv.org/abs/2405.05068} {\bibinfo {title} {Chemistry beyond exact solutions on a quantum-centric supercomputer}} (\bibinfo {year} {2024}),\ \Eprint {https://arxiv.org/abs/2405.05068} {arXiv:2405.05068 [quant-ph]} \BibitemShut {NoStop}%
\bibitem [{\citenamefont {Greene-Diniz}\ \emph {et~al.}(2024)\citenamefont {Greene-Diniz}, \citenamefont {Self}, \citenamefont {Krompiec}, \citenamefont {Coopmans}, \citenamefont {Benedetti}, \citenamefont {Ramo},\ and\ \citenamefont {Rosenkranz}}]{Greenediniz2024}%
  \BibitemOpen
  \bibfield  {author} {\bibinfo {author} {\bibfnamefont {G.}~\bibnamefont {Greene-Diniz}}, \bibinfo {author} {\bibfnamefont {C.~N.}\ \bibnamefont {Self}}, \bibinfo {author} {\bibfnamefont {M.}~\bibnamefont {Krompiec}}, \bibinfo {author} {\bibfnamefont {L.}~\bibnamefont {Coopmans}}, \bibinfo {author} {\bibfnamefont {M.}~\bibnamefont {Benedetti}}, \bibinfo {author} {\bibfnamefont {D.~M.}\ \bibnamefont {Ramo}},\ and\ \bibinfo {author} {\bibfnamefont {M.}~\bibnamefont {Rosenkranz}},\ }\href {https://arxiv.org/abs/2409.15908} {\bibinfo {title} {Measuring correlation and entanglement between molecular orbitals on a trapped-ion quantum computer}} (\bibinfo {year} {2024}),\ \Eprint {https://arxiv.org/abs/2409.15908} {arXiv:2409.15908 [quant-ph]} \BibitemShut {NoStop}%
\bibitem [{\citenamefont {Gerasimov}\ \emph {et~al.}(2022)\citenamefont {Gerasimov}, \citenamefont {Yusupov}, \citenamefont {Moiseevsky}, \citenamefont {Vybornyi}, \citenamefont {Tikhonov}, \citenamefont {Kulik}, \citenamefont {Straupe}, \citenamefont {Sukenik},\ and\ \citenamefont {Kupriyanov}}]{Gerasimov2022}%
  \BibitemOpen
  \bibfield  {author} {\bibinfo {author} {\bibfnamefont {L.~V.}\ \bibnamefont {Gerasimov}}, \bibinfo {author} {\bibfnamefont {R.~R.}\ \bibnamefont {Yusupov}}, \bibinfo {author} {\bibfnamefont {A.~D.}\ \bibnamefont {Moiseevsky}}, \bibinfo {author} {\bibfnamefont {I.}~\bibnamefont {Vybornyi}}, \bibinfo {author} {\bibfnamefont {K.~S.}\ \bibnamefont {Tikhonov}}, \bibinfo {author} {\bibfnamefont {S.~P.}\ \bibnamefont {Kulik}}, \bibinfo {author} {\bibfnamefont {S.~S.}\ \bibnamefont {Straupe}}, \bibinfo {author} {\bibfnamefont {C.~I.}\ \bibnamefont {Sukenik}},\ and\ \bibinfo {author} {\bibfnamefont {D.~V.}\ \bibnamefont {Kupriyanov}},\ }\bibfield  {title} {\bibinfo {title} {Coupled dynamics of spin qubits in optical dipole microtraps: Application to the error analysis of a rydberg-blockade gate},\ }\href {https://doi.org/10.1103/PhysRevA.106.042410} {\bibfield  {journal} {\bibinfo  {journal} {Phys. Rev. A}\ }\textbf {\bibinfo {volume} {106}},\ \bibinfo {pages} {042410} (\bibinfo {year} {2022})}\BibitemShut {NoStop}%
\bibitem [{\citenamefont {Vybornyi}\ \emph {et~al.}(2024)\citenamefont {Vybornyi}, \citenamefont {Gerasimov}, \citenamefont {Kupriyanov}, \citenamefont {Straupe},\ and\ \citenamefont {Tikhonov}}]{Vybornyi2024}%
  \BibitemOpen
  \bibfield  {author} {\bibinfo {author} {\bibfnamefont {I.}~\bibnamefont {Vybornyi}}, \bibinfo {author} {\bibfnamefont {L.~V.}\ \bibnamefont {Gerasimov}}, \bibinfo {author} {\bibfnamefont {D.~V.}\ \bibnamefont {Kupriyanov}}, \bibinfo {author} {\bibfnamefont {S.~S.}\ \bibnamefont {Straupe}},\ and\ \bibinfo {author} {\bibfnamefont {K.~S.}\ \bibnamefont {Tikhonov}},\ }\bibfield  {title} {\bibinfo {title} {Influence of the interaction geometry on the fidelity of the two-qubit rydberg blockade gate},\ }\href {https://doi.org/10.1364/JOSAB.504629} {\bibfield  {journal} {\bibinfo  {journal} {J. Opt. Soc. Am. B}\ }\textbf {\bibinfo {volume} {41}},\ \bibinfo {pages} {134} (\bibinfo {year} {2024})}\BibitemShut {NoStop}%
\bibitem [{\citenamefont {Graham}\ \emph {et~al.}(2023)\citenamefont {Graham}, \citenamefont {Phuttitarn}, \citenamefont {Chinnarasu}, \citenamefont {Song}, \citenamefont {Poole}, \citenamefont {Jooya}, \citenamefont {Scott}, \citenamefont {Scott}, \citenamefont {Eichler},\ and\ \citenamefont {Saffman}}]{Saffman2023}%
  \BibitemOpen
  \bibfield  {author} {\bibinfo {author} {\bibfnamefont {T.}~\bibnamefont {Graham}}, \bibinfo {author} {\bibfnamefont {L.}~\bibnamefont {Phuttitarn}}, \bibinfo {author} {\bibfnamefont {R.}~\bibnamefont {Chinnarasu}}, \bibinfo {author} {\bibfnamefont {Y.}~\bibnamefont {Song}}, \bibinfo {author} {\bibfnamefont {C.}~\bibnamefont {Poole}}, \bibinfo {author} {\bibfnamefont {K.}~\bibnamefont {Jooya}}, \bibinfo {author} {\bibfnamefont {J.}~\bibnamefont {Scott}}, \bibinfo {author} {\bibfnamefont {A.}~\bibnamefont {Scott}}, \bibinfo {author} {\bibfnamefont {P.}~\bibnamefont {Eichler}},\ and\ \bibinfo {author} {\bibfnamefont {M.}~\bibnamefont {Saffman}},\ }\bibfield  {title} {\bibinfo {title} {Midcircuit measurements on a single-species neutral alkali atom quantum processor},\ }\bibfield  {journal} {\bibinfo  {journal} {Physical Review X}\ }\textbf {\bibinfo {volume} {13}},\ \href {https://doi.org/10.1103/physrevx.13.041051} {10.1103/physrevx.13.041051} (\bibinfo {year} {2023})\BibitemShut {NoStop}%
\bibitem [{\citenamefont {Wagner}\ \emph {et~al.}(2024)\citenamefont {Wagner}, \citenamefont {Poole}, \citenamefont {Graham},\ and\ \citenamefont {Saffman}}]{Saffman2024a}%
  \BibitemOpen
  \bibfield  {author} {\bibinfo {author} {\bibfnamefont {N.}~\bibnamefont {Wagner}}, \bibinfo {author} {\bibfnamefont {C.}~\bibnamefont {Poole}}, \bibinfo {author} {\bibfnamefont {T.~M.}\ \bibnamefont {Graham}},\ and\ \bibinfo {author} {\bibfnamefont {M.}~\bibnamefont {Saffman}},\ }\bibfield  {title} {\bibinfo {title} {Benchmarking a neutral-atom quantum computer},\ }\bibfield  {journal} {\bibinfo  {journal} {International Journal of Quantum Information}\ }\textbf {\bibinfo {volume} {22}},\ \href {https://doi.org/10.1142/s0219749924500011} {10.1142/s0219749924500011} (\bibinfo {year} {2024})\BibitemShut {NoStop}%
\bibitem [{\citenamefont {Poole}\ \emph {et~al.}(2024)\citenamefont {Poole}, \citenamefont {Graham}, \citenamefont {Perlin}, \citenamefont {Otten},\ and\ \citenamefont {Saffman}}]{Saffman2024b}%
  \BibitemOpen
  \bibfield  {author} {\bibinfo {author} {\bibfnamefont {C.}~\bibnamefont {Poole}}, \bibinfo {author} {\bibfnamefont {T.~M.}\ \bibnamefont {Graham}}, \bibinfo {author} {\bibfnamefont {M.~A.}\ \bibnamefont {Perlin}}, \bibinfo {author} {\bibfnamefont {M.}~\bibnamefont {Otten}},\ and\ \bibinfo {author} {\bibfnamefont {M.}~\bibnamefont {Saffman}},\ }\href {https://arxiv.org/abs/2404.18809} {\bibinfo {title} {Architecture for fast implementation of qldpc codes with optimized rydberg gates}} (\bibinfo {year} {2024}),\ \Eprint {https://arxiv.org/abs/2404.18809} {arXiv:2404.18809 [quant-ph]} \BibitemShut {NoStop}%
\bibitem [{\citenamefont {Daley}\ \emph {et~al.}(2022)\citenamefont {Daley}, \citenamefont {Bloch}, \citenamefont {Kokail}, \citenamefont {Flannigan}, \citenamefont {Pearson}, \citenamefont {Troyer},\ and\ \citenamefont {Zoller}}]{Zoller2022}%
  \BibitemOpen
  \bibfield  {author} {\bibinfo {author} {\bibfnamefont {A.~J.}\ \bibnamefont {Daley}}, \bibinfo {author} {\bibfnamefont {I.}~\bibnamefont {Bloch}}, \bibinfo {author} {\bibfnamefont {C.}~\bibnamefont {Kokail}}, \bibinfo {author} {\bibfnamefont {S.}~\bibnamefont {Flannigan}}, \bibinfo {author} {\bibfnamefont {N.}~\bibnamefont {Pearson}}, \bibinfo {author} {\bibfnamefont {M.}~\bibnamefont {Troyer}},\ and\ \bibinfo {author} {\bibfnamefont {P.}~\bibnamefont {Zoller}},\ }\bibfield  {title} {\bibinfo {title} {Practical quantum advantage in quantum simulation},\ }\href {https://doi.org/10.1038/s41586-022-04940-6} {\bibfield  {journal} {\bibinfo  {journal} {Nature}\ }\textbf {\bibinfo {volume} {607}},\ \bibinfo {pages} {667–676} (\bibinfo {year} {2022})}\BibitemShut {NoStop}%
\bibitem [{\citenamefont {Argüello-Luengo}\ \emph {et~al.}(2019)\citenamefont {Argüello-Luengo}, \citenamefont {González-Tudela}, \citenamefont {Shi}, \citenamefont {Zoller},\ and\ \citenamefont {Cirac}}]{Cirac2019}%
  \BibitemOpen
  \bibfield  {author} {\bibinfo {author} {\bibfnamefont {J.}~\bibnamefont {Argüello-Luengo}}, \bibinfo {author} {\bibfnamefont {A.}~\bibnamefont {González-Tudela}}, \bibinfo {author} {\bibfnamefont {T.}~\bibnamefont {Shi}}, \bibinfo {author} {\bibfnamefont {P.}~\bibnamefont {Zoller}},\ and\ \bibinfo {author} {\bibfnamefont {J.~I.}\ \bibnamefont {Cirac}},\ }\bibfield  {title} {\bibinfo {title} {Analogue quantum chemistry simulation},\ }\href {https://doi.org/10.1038/s41586-019-1614-4} {\bibfield  {journal} {\bibinfo  {journal} {Nature}\ }\textbf {\bibinfo {volume} {574}},\ \bibinfo {pages} {215–218} (\bibinfo {year} {2019})}\BibitemShut {NoStop}%
\bibitem [{\citenamefont {Argüello-Luengo}\ \emph {et~al.}(2022)\citenamefont {Argüello-Luengo}, \citenamefont {González-Tudela},\ and\ \citenamefont {González-Cuadra}}]{Arguello_Luengo2022}%
  \BibitemOpen
  \bibfield  {author} {\bibinfo {author} {\bibfnamefont {J.}~\bibnamefont {Argüello-Luengo}}, \bibinfo {author} {\bibfnamefont {A.}~\bibnamefont {González-Tudela}},\ and\ \bibinfo {author} {\bibfnamefont {D.}~\bibnamefont {González-Cuadra}},\ }\bibfield  {title} {\bibinfo {title} {Tuning long-range fermion-mediated interactions in cold-atom quantum simulators},\ }\bibfield  {journal} {\bibinfo  {journal} {Physical Review Letters}\ }\textbf {\bibinfo {volume} {129}},\ \href {https://doi.org/10.1103/physrevlett.129.083401} {10.1103/physrevlett.129.083401} (\bibinfo {year} {2022})\BibitemShut {NoStop}%
\bibitem [{\citenamefont {Kaufman}\ \emph {et~al.}(2015)\citenamefont {Kaufman}, \citenamefont {Lester}, \citenamefont {Foss-Feig}, \citenamefont {Wall}, \citenamefont {Rey},\ and\ \citenamefont {Regal}}]{Kaufman2015}%
  \BibitemOpen
  \bibfield  {author} {\bibinfo {author} {\bibfnamefont {A.~M.}\ \bibnamefont {Kaufman}}, \bibinfo {author} {\bibfnamefont {B.~J.}\ \bibnamefont {Lester}}, \bibinfo {author} {\bibfnamefont {M.}~\bibnamefont {Foss-Feig}}, \bibinfo {author} {\bibfnamefont {M.~L.}\ \bibnamefont {Wall}}, \bibinfo {author} {\bibfnamefont {A.~M.}\ \bibnamefont {Rey}},\ and\ \bibinfo {author} {\bibfnamefont {C.~A.}\ \bibnamefont {Regal}},\ }\bibfield  {title} {\bibinfo {title} {Entangling two transportable neutral atoms via local spin exchange},\ }\bibfield  {journal} {\bibinfo  {journal} {Nature}\ }\textbf {\bibinfo {volume} {527}},\ \href {https://doi.org/10.1038/nature16073} {10.1038/nature16073} (\bibinfo {year} {2015})\BibitemShut {NoStop}%
\bibitem [{\citenamefont {Brandt}\ \emph {et~al.}(2018)\citenamefont {Brandt}, \citenamefont {Yannouleas},\ and\ \citenamefont {Landman}}]{Brandt2018}%
  \BibitemOpen
  \bibfield  {author} {\bibinfo {author} {\bibfnamefont {B.~B.}\ \bibnamefont {Brandt}}, \bibinfo {author} {\bibfnamefont {C.}~\bibnamefont {Yannouleas}},\ and\ \bibinfo {author} {\bibfnamefont {U.}~\bibnamefont {Landman}},\ }\bibfield  {title} {\bibinfo {title} {Interatomic interaction effects on second-order momentum correlations and hong-ou-mandel interference of double-well-trapped ultracold fermionic atoms},\ }\bibfield  {journal} {\bibinfo  {journal} {Physical Review A}\ }\textbf {\bibinfo {volume} {97}},\ \href {https://doi.org/10.1103/physreva.97.053601} {10.1103/physreva.97.053601} (\bibinfo {year} {2018})\BibitemShut {NoStop}%
\bibitem [{\citenamefont {Kaufman}\ \emph {et~al.}(2018)\citenamefont {Kaufman}, \citenamefont {Tichy}, \citenamefont {Mintert}, \citenamefont {Rey},\ and\ \citenamefont {Regal}}]{Regal2018}%
  \BibitemOpen
  \bibfield  {author} {\bibinfo {author} {\bibfnamefont {A.~M.}\ \bibnamefont {Kaufman}}, \bibinfo {author} {\bibfnamefont {M.~C.}\ \bibnamefont {Tichy}}, \bibinfo {author} {\bibfnamefont {F.}~\bibnamefont {Mintert}}, \bibinfo {author} {\bibfnamefont {A.~M.}\ \bibnamefont {Rey}},\ and\ \bibinfo {author} {\bibfnamefont {C.~A.}\ \bibnamefont {Regal}},\ }\bibinfo {title} {The hong–ou–mandel effect with atoms},\ in\ \href {https://doi.org/10.1016/bs.aamop.2018.03.003} {\emph {\bibinfo {booktitle} {Advances In Atomic, Molecular, and Optical Physics}}}\ (\bibinfo  {publisher} {Elsevier},\ \bibinfo {year} {2018})\ p.\ \bibinfo {pages} {377–427}\BibitemShut {NoStop}%
\bibitem [{\citenamefont {Aspect}(2020)}]{Aspect2020}%
  \BibitemOpen
  \bibfield  {author} {\bibinfo {author} {\bibfnamefont {A.}~\bibnamefont {Aspect}},\ }\href {https://arxiv.org/abs/2005.08239} {\bibinfo {title} {Hanburry brown and twiss, hong ou and mandel effects and other landmarks in quantum optics: from photons to atoms}} (\bibinfo {year} {2020}),\ \Eprint {https://arxiv.org/abs/2005.08239} {arXiv:2005.08239 [quant-ph]} \BibitemShut {NoStop}%
\bibitem [{\citenamefont {Mivehvar}\ \emph {et~al.}(2017)\citenamefont {Mivehvar}, \citenamefont {Ritsch},\ and\ \citenamefont {Piazza}}]{Ritsch2017}%
  \BibitemOpen
  \bibfield  {author} {\bibinfo {author} {\bibfnamefont {F.}~\bibnamefont {Mivehvar}}, \bibinfo {author} {\bibfnamefont {H.}~\bibnamefont {Ritsch}},\ and\ \bibinfo {author} {\bibfnamefont {F.}~\bibnamefont {Piazza}},\ }\bibfield  {title} {\bibinfo {title} {Superradiant topological peierls insulator inside an optical cavity},\ }\href {https://doi.org/10.1103/PhysRevLett.118.073602} {\bibfield  {journal} {\bibinfo  {journal} {Phys. Rev. Lett.}\ }\textbf {\bibinfo {volume} {118}},\ \bibinfo {pages} {073602} (\bibinfo {year} {2017})}\BibitemShut {NoStop}%
\bibitem [{\citenamefont {Colella}\ \emph {et~al.}(2019)\citenamefont {Colella}, \citenamefont {Ostermann}, \citenamefont {Niedenzu}, \citenamefont {Mivehvar},\ and\ \citenamefont {Ritsch}}]{Ritsch2019}%
  \BibitemOpen
  \bibfield  {author} {\bibinfo {author} {\bibfnamefont {E.}~\bibnamefont {Colella}}, \bibinfo {author} {\bibfnamefont {S.}~\bibnamefont {Ostermann}}, \bibinfo {author} {\bibfnamefont {W.}~\bibnamefont {Niedenzu}}, \bibinfo {author} {\bibfnamefont {F.}~\bibnamefont {Mivehvar}},\ and\ \bibinfo {author} {\bibfnamefont {H.}~\bibnamefont {Ritsch}},\ }\bibfield  {title} {\bibinfo {title} {Antiferromagnetic self-ordering of a fermi gas in a ring cavity},\ }\href {https://doi.org/10.1088/1367-2630/ab151e} {\bibfield  {journal} {\bibinfo  {journal} {New Journal of Physics}\ }\textbf {\bibinfo {volume} {21}},\ \bibinfo {pages} {043019} (\bibinfo {year} {2019})}\BibitemShut {NoStop}%
\bibitem [{\citenamefont {Goldberger}\ and\ \citenamefont {Watson}(1964)}]{GoldWat1964}%
  \BibitemOpen
  \bibfield  {author} {\bibinfo {author} {\bibfnamefont {M.~L.}\ \bibnamefont {Goldberger}}\ and\ \bibinfo {author} {\bibfnamefont {K.~M.}\ \bibnamefont {Watson}},\ }\href {https://www.bibsonomy.org/bibtex/2e7ca430ae6d854e9fc55257e34c0ad03/bronckobuster} {\emph {\bibinfo {title} {Collision Theory}}}\ (\bibinfo  {publisher} {Wiley},\ \bibinfo {address} {New York},\ \bibinfo {year} {1964})\BibitemShut {NoStop}%
\bibitem [{\citenamefont {Zahradník}\ and\ \citenamefont {Polák}(1980)}]{ZahradníkPolak1980}%
  \BibitemOpen
  \bibfield  {author} {\bibinfo {author} {\bibfnamefont {R.}~\bibnamefont {Zahradník}}\ and\ \bibinfo {author} {\bibfnamefont {R.}~\bibnamefont {Polák}},\ }\href {https://doi.org/10.1007/978-1-4613-3921-2} {\emph {\bibinfo {title} {Elements of Quantum Chemistry}}}\ (\bibinfo  {publisher} {Springer New York, NY},\ \bibinfo {year} {1980})\BibitemShut {NoStop}%
\bibitem [{\citenamefont {Roothaan}(1951)}]{Roothaan1951}%
  \BibitemOpen
  \bibfield  {author} {\bibinfo {author} {\bibfnamefont {C.~C.~J.}\ \bibnamefont {Roothaan}},\ }\bibfield  {title} {\bibinfo {title} {New developments in molecular orbital theory},\ }\href {https://doi.org/10.1103/RevModPhys.23.69} {\bibfield  {journal} {\bibinfo  {journal} {Rev. Mod. Phys.}\ }\textbf {\bibinfo {volume} {23}},\ \bibinfo {pages} {69} (\bibinfo {year} {1951})}\BibitemShut {NoStop}%
\bibitem [{\citenamefont {Landau}\ and\ \citenamefont {Lifshitz}(1981)}]{LandauLifshitzIII}%
  \BibitemOpen
  \bibfield  {author} {\bibinfo {author} {\bibfnamefont {L.~D.}\ \bibnamefont {Landau}}\ and\ \bibinfo {author} {\bibfnamefont {E.~M.}\ \bibnamefont {Lifshitz}},\ }\href {https://books.google.ru/books?id=SvdoN3k8EysC} {\emph {\bibinfo {title} {Quantum Mechanics: Non-Relativistic Theory}}},\ Course of theoretical physics\ (\bibinfo  {publisher} {Butterworth-Heinemann},\ \bibinfo {year} {1981})\BibitemShut {NoStop}%
\bibitem [{\citenamefont {Kanno}\ \emph {et~al.}(2010)\citenamefont {Kanno}, \citenamefont {Kono}, \citenamefont {Fujimura},\ and\ \citenamefont {Lin}}]{Kanno2010}%
  \BibitemOpen
  \bibfield  {author} {\bibinfo {author} {\bibfnamefont {M.}~\bibnamefont {Kanno}}, \bibinfo {author} {\bibfnamefont {H.}~\bibnamefont {Kono}}, \bibinfo {author} {\bibfnamefont {Y.}~\bibnamefont {Fujimura}},\ and\ \bibinfo {author} {\bibfnamefont {S.~H.}\ \bibnamefont {Lin}},\ }\bibfield  {title} {\bibinfo {title} {Nonadiabatic response model of laser-induced ultrafast $\ensuremath{\pi}$-electron rotations in chiral aromatic molecules},\ }\href {https://doi.org/10.1103/PhysRevLett.104.108302} {\bibfield  {journal} {\bibinfo  {journal} {Phys. Rev. Lett.}\ }\textbf {\bibinfo {volume} {104}},\ \bibinfo {pages} {108302} (\bibinfo {year} {2010})}\BibitemShut {NoStop}%
\bibitem [{\citenamefont {Messiah}(1961)}]{Messiah1961}%
  \BibitemOpen
  \bibfield  {author} {\bibinfo {author} {\bibfnamefont {A.}~\bibnamefont {Messiah}},\ }\href {https://www.bibsonomy.org/bibtex/2d4a990fdc8e76985bfa727b633aa8c1d/mcclung} {\emph {\bibinfo {title} {Quantum Mechanics Volume II}}}\ (\bibinfo  {publisher} {Elsevier Science B.V.},\ \bibinfo {year} {1961})\BibitemShut {NoStop}%
\bibitem [{\citenamefont {Varshalovich}\ \emph {et~al.}(1988)\citenamefont {Varshalovich}, \citenamefont {Moskalev},\ and\ \citenamefont {Khersonskii}}]{Varshalovich1988}%
  \BibitemOpen
  \bibfield  {author} {\bibinfo {author} {\bibfnamefont {D.~A.}\ \bibnamefont {Varshalovich}}, \bibinfo {author} {\bibfnamefont {A.~N.}\ \bibnamefont {Moskalev}},\ and\ \bibinfo {author} {\bibfnamefont {V.~K.}\ \bibnamefont {Khersonskii}},\ }\href {https://doi.org/10.1142/0270} {\emph {\bibinfo {title} {Quantum Theory of Angular Momentum}}}\ (\bibinfo  {publisher} {World Scientific Publishing Company},\ \bibinfo {year} {1988})\BibitemShut {NoStop}%
\end{thebibliography}%

\clearpage
\onecolumngrid
\pagestyle{empty}   

\begin{center}
{\large\bfseries Erratum: Spin versus position conjugation in quantum simulations with atoms:\\ Application to quantum chemistry [Phys. Rev. A 111, 062823 (2025)]}

\vspace{1.5em}

N.A. Moroz,
K.S. Tikhonov,
L.V. Gerasimov,
A.D. Manukhova,
I.B. Bobrov,
S.S. Straupe,
D.V. Kupriyanov

\vspace{0.3em}


\vspace{1em}
\end{center}

\title{Erratum: Spin versus position conjugation in quantum simulations with atoms:\\ Application to quantum chemistry [Phys. Rev. A 111, 062823 (2025)]}%



\maketitle


\noindent In Appendix A Eq.~(A14) contains incorrect sign in its last line. As a consequence we withdraw Eq.~(A15) and correct the related equations (B5) as
\begin{equation}
\rho^{(\mathrm{int})}(1,2,3)=-\frac{\sqrt{3}}{2}\sum_{\circlearrowleft}\left[\Phi^{(0;1/2)}(i;j,k)\cdot\Phi^{(1;1/2)\ast}(k;i,j)-\Phi^{(0;1/2)}(k;i,j)\cdot\Phi^{(1;1/2)\ast}(i;j,k)\right]%
\ \ \ \ \ \ \ \ \ \ \ \ \ \ (\mathrm{B5})%
\nonumber\\%
\end{equation}
and (B10) as
\begin{equation}
\rho^{(\mathrm{int})}(1,2,3,4)%
=-\frac{\sqrt{3}}{2}\sum_{\circlearrowleft}\left[\Phi^{(0,0;0)}(i,j;k,4)\cdot\Phi^{(1,1;0)\ast}(k,i;j,4)-\Phi^{(0,0;0)}(k,i;j,4)\cdot\Phi^{(1,1;0)\ast}(i,j;k,4)\right]%
\ \ \ \  (\mathrm{B10})%
\nonumber\\%
\end{equation}
where $\circlearrowleft$ denotes the sum over three possible cyclic permutations $(1,2,3)\to (i,j,k)$. 

Being set by the products of mutually orthogonal single-particle orbitals, the complete spin-position wavefunctions, belonging to different spin coupling schemes, are orthogonal if the particles occupy different orbitals. Otherwise, both the considered coupling schemes lead to identical wavefunctions and coincidence in the spatial distributions in accordance with the calculations presented in the main text. 

This correction has no action on our principle results, discussed in the main text.

\end{document}